\documentclass[nofootinbib,prd,twocolumn,showpacs,showkeys,preprintnumbers]{revtex4-1}
\usepackage{hyperref,amssymb,amsmath,mathrsfs,bm,graphicx}
\usepackage{multirow}
\usepackage{array}
\usepackage{float}
\usepackage{placeins}
\usepackage[dvipsnames]{xcolor}
\begin{document}
\title{A static and spherically symmetric hairy black hole in the framework of the Gravitational Decoupling.}
\author{R. Avalos}
\email{ravalos@usfq.edu.ec}
\affiliation{Departamento de F\'isica, Colegio de Ciencias e Ingenier\'ia, Universidad San Francisco de Quito,  Quito 170901, Ecuador.}

\author{Pedro Bargue\~no }
\email{pedro.bargueno@ua.es}
\address{Departamento de F\'{\i}sica Aplicada, Universidad de Alicante, Campus de San Vicente del Raspeig, E-03690 Alicante, Spain}

\author{E. Contreras }
\email{econtreras@usfq.edu.ec}
\affiliation{Departamento de F\'isica, Colegio de Ciencias e Ingenier\'ia, Universidad San Francisco de Quito,  Quito 170901, Ecuador.\\}

\begin{abstract}
In this work we construct a static and spherically symmetric black hole geometry supported by a family of generic mono-parametric sources thorough the Gravitational Decoupling. The parameter characterizing the matter sector can be interpreted as hair which cannot be associated to any global charge.  Although the solution is constructed by demanding the weak energy condition, we find that the resulting matter sector satisfies all the energy conditions at and outside the horizon. We study the effect of the hair on the periastron advance and the gravitational lensing of the black hole. We estimate the best WKB order to compute the quasinormal frequencies for scalar, vector and tensor perturbation fields.  
\end{abstract}

\maketitle

\section{Introduction}
The uniqueness theorems allow to conjecture that the final stage of a gravitational collapse should be a Kerr--Newmann black hole (BH) independently of the existent matter before the collapse. In other words, the conjecture states that these solutions can only be characterized by 
{\it global charges}, namely the mass, angular momentum and electric charge, all of these asymptotically measured quantities satisfying a Gauss law  \cite{Herdeiro:2015waa}. This {\it non--hair conjecture}, as it is broadly known, leads to conclude that, two black holes with same mass, charge and angular momentum are indistinguishable independent of the manner they were formed. However, there are well-known cases of BH's solutions supported with non--linear matter sources (Einstein--Skyrme \cite{Luckock:1986tr,Glendenning:1988qy,Bizon:1992gb} and Einstein-Yang--Mills \cite{Bartnik:1988am,Bizon:1990sr} models, for example) which have served as counter examples of the stated by the non--hair conjecture \cite{Bizon:1994dh,Bekenstein:1996pn}. In this regard, some effort have been made to reformulate the non--hair conjecture to include the counter examples in a formal way. However, the conclusion seems always to be the same: {\it  there is no universally valid formulation of the no-hair conjecture} \cite{Bizon:1994dh}. In this regard, it is always possible to find suitable black hole solutions with hairs that cannot be associated to global charges. More precisely, solutions
that are indistinguishable from a Kerr--Newmann BH by an asymptotic observer.

It is our main goal here to construct a hairy black hole with a generic matter sector through the Gravitational Decoupling (GD) approach \cite{Ovalle:2017fgl,Ovalle:2019qyi} (see \cite{Fernandes-Silva:2018abr,daRocha:2019pla,Heras:2019ibr,daRocha:2020rda,daRocha:2020jdj,Tello-Ortiz:2020euy,daRocha:2020gee,Meert:2020sqv,Tello-Ortiz:2021kxg,Maurya:2021huv,Azmat:2021kmv,Maurya:2021zvb,Ovalle:2017wqi,Gabbanelli:2018bhs,Heras:2018cpz,Estrada:2018zbh,Morales:2018urp,Estrada:2018vrl,Ovalle:2018umz,Ovalle:2018ans,Gabbanelli:2019txr,Estrada:2019aeh,Ovalle:2019lbs,Casadio:2019usg,Singh:2019ktp,Maurya:2019noq,Tello-Ortiz:2020ydf,Maurya:2020rny,Rincon:2020izv,Maurya:2020gjw,Zubair:2020lna,Sharif:2020rlt,Ovalle:2020kpd,Ovalle:2020fuo,Estrada:2020ptc,Maurya:2020djz,Meert:2021khi,Maurya:2021aio,Azmat:2021qig,Islam:2021dyk,Afrin:2021imp,Ovalle:2021jzf,Ama-Tul-Mughani:2021ewd,daRocha:2021aww,Maurya:2021qye,Carrasco-Hidalgo:2021dyg,Sultana:2021cvq,daRocha:2021sqd,Maurya:2021yhc,Omwoyo:2021uah,Afrin:2021ggx,Ovalle:2022eqb,Andrade,Dayanandan:2021odo,Contreras:2021yxe} for applications of GD in standard general relativity. For applications in higher dimensions see \cite{Maurya:2022brt,Maurya:2022uqu}, for example). Of course, this is not the first time that the GD is used as a tool to construct {\it hairy }black holes. For example, in Ref. \cite{Ovalle:2018umz}, the authors implemented the GD by the Minimal Geometric Deformation (MGD) to construct BH's supported with a generic source satisfying different equations of state. Then, the solution in \cite{Ovalle:2018umz} satisfying a linear equation of state was used to model a BH solution with a scalar hair in \cite{Ovalle:2018ans}. More recently, in \cite{Ovalle:2020kpd} the authors found a hairy solution regular on and outside the horizon satisfying the strong (SEC) and the dominant energy conditions (DEC) everywhere outside the horizon. In Ref. \cite{Cavalcanti:2022adb}, the thermodynamics of \cite{Ovalle:2020kpd} has been analyzed in detail and in Ref. \cite{Zhang:2022niv} a new solution of hairy black hole in asymptotically AdS space--time have been found. In contrast to the previous work mentioned above, our construction is based on the less restrictive weak energy condition (WEC) on and outside the horizon which implies that the local energy density cannot be negative for all observers.

This work is organized as follows. In the next, section we introduce the GD approach. Next, in section \ref{HBH} we apply the GD to construct a BH satisfying the WEC. Section \ref{GM} is devoted to the analysis of the geodesics. In particular, we study how the periastron advance and the gravitational lensing are affected by the hair of our model. In section (\ref{QNM}) we compute the quasinormal modes (QNM) of the solution for scalar, vector and tensor perturbing fields. In  particular, we first estimate the best order of the WKB semi--analytical formula and then, we compute the imaginary part of the QNM to the best order. In the last section we conclude our work.

\section{Gravitational decoupling} \label{GD}

Les us consider the Einstein field equations 
\begin{eqnarray}\label{EE}
G_{\mu \nu} \equiv R_{\mu \nu}-\frac{1}{2} R g_{\mu \nu}=k^2 \Tilde{T}_{\mu \nu},
\end{eqnarray}
where $k^2=\frac{8 \pi G}{c^4}$ and $\Tilde{T}_{\mu \nu}$ is the total energy momentum tensor containing two contributions,
\begin{eqnarray} \label{desacople}
\Tilde{T}_{\mu \nu} = T_{\mu \nu} + \theta _{\mu \nu},
\end{eqnarray}
with $T_{\mu \nu}$ is associated with some known solution which will be used as a seed, while $\theta_{\mu \nu}$ represents new contributions to the fields or the gravitational sector. The Einstein tensor accomplishes the Bianchi identity, which means that the source must be covariantly conserved,
\begin{eqnarray}
\nabla_{\mu} \Tilde{T}^{\mu \nu}=0.
\end{eqnarray}

Considering a spherically symmetric and static system, the metric can me written as
\begin{eqnarray} \label{sphericmetric}
ds^{2}=e^{\nu(r)} dt^{2} -e^{\lambda (r)} dr^{2}-r^{2}d\Omega^{2},
\end{eqnarray}
where $\nu=\nu(r)$ and $\lambda=\lambda(r)$ are functions dependent only of the radius $r$ and $d\Omega^{2}=d\theta^2+\sin^2 \theta d\phi^2$. Then, the Einstein equations \eqref{EE} read
\begin{eqnarray}
k^2 \big( T_0^0+\theta_0^0 \big)&=&\frac{1}{r^2}-e^{-\lambda}\bigg(\frac{1}{r^2}-\frac{\lambda '}{r} \bigg),\label{T0}\\
k^2 \big( T_1^1+\theta_0^0 \big)&=&\frac{1}{r^2}-e^{-\lambda}\bigg(\frac{1}{r^2}+\frac{\nu '}{r} \bigg),\label{T1}\\
k^2 \big( T_2^2+\theta_0^0 \big)&=&-\frac{e^{-\lambda}}{4}\bigg(2\nu '' +\nu'^2-\lambda' \nu' +2\frac{\nu'-\lambda'}{r} \bigg),\label{T2}
\end{eqnarray}
where $f'\equiv \partial_r f$ and $\Tilde{T}_2^2=\Tilde{T}_3^3$ due to the spherical symmetry. We can define an effective matter sector by using Eqs. \eqref{T0}-\eqref{T2}, from where
\begin{eqnarray}\label{rhoeff}
\Tilde{\rho}=T_0^0+\theta_0^0,
\end{eqnarray}
the effective radial pressure
\begin{eqnarray}\label{preff}
\Tilde{p}_r=-T_1^1-\theta_1^1,
\end{eqnarray}
and the effective tangential pressure
\begin{eqnarray}\label{pteff}
\Tilde{p}_t=-T_2^2-\theta_2^2.
\end{eqnarray}

Next, let us consider a solution for Eq. \eqref{EE} for the seed source $T_{\mu \nu}$, which means not considering the term $\theta_{\mu \nu}$. We can write a metric
\begin{eqnarray}\label{metricseed}
ds^{2}=e^{\xi(r)} dt^{2} -e^{\mu (r)} dr^{2}-r^{2}d\Omega^{2},
\end{eqnarray}
where
\begin{eqnarray}
e^{-\mu(r)}\equiv1-\frac{k^2}{r} \int_0^r x^2 T_0^0(x)dx=1-\frac{2m(r)}{r}
\end{eqnarray}
contains the mass function $m=m(r)$. We can add the source $\theta_{\mu \nu}$ using the extended geometric deformation (EGD) on the seed metric, obtaining
\begin{eqnarray}
\xi &\rightarrow& \nu=\xi+ g, \label{EGD1}\\
e^{-\nu} &\rightarrow& e^{-\lambda}=e^{-\mu}+ f, \label{EGD2}
\end{eqnarray}
where $f$ and $g$ are the geometric deformations corresponding to the radial and temporal components, respectively. Eqs. \eqref{EGD1} and \eqref{EGD2} can be used to separate the Einstein equations \eqref{T0}-\eqref{T2} into two different sets. The first one is sourced by the energy momentum tensor, $T_{\mu \nu}$, which is the seed, and reads
\begin{eqnarray}
k^2 T_0^0&=&\frac{1}{r^2}-e^{-\mu}\bigg(\frac{1}{r^2}-\frac{\mu '}{r} \bigg),\label{T0seed}\\
k^2 T_1^1&=&\frac{1}{r^2}-e^{-\mu}\bigg(\frac{1}{r^2}+\frac{\xi '}{r} \bigg),\label{T1seed}\\
k^2 T_2^2&=&-\frac{e^{-\mu}}{4}\bigg(2\xi '' +\xi'^2-\mu' \xi' +2\frac{\xi'-\mu'}{r} \bigg),\label{T2seed}
\end{eqnarray}
The second one contains the source, $\theta_{\mu \nu}$, and reads
\begin{eqnarray}
k^2 \theta_0^0&=&- \frac{f}{r^2}-\frac{f'}{r},\label{T0source}\\
k^2 \theta_1^1&=&- Z_1 - f \bigg( \frac{1}{r^2}+\frac{\nu'}{r} \bigg),\label{T1source}\\
k^2 \theta_2^2&=&- Z_2 - \frac{f}{4} \bigg( 2\nu''+
\nu'^2+2\frac{\nu'}{r}\bigg) \nonumber \\
&&- \frac{f'}{4}\bigg( \nu' +\frac{2}{r} \bigg),\label{T2source}
\end{eqnarray}
where
\begin{eqnarray}
Z_1&=&\frac{e^{-\mu} g'}{r}, \\
Z_2&=&\frac{e^{-\mu}}{4} \bigg( 2g''+ g'^2 +2\frac{g'}{r}+2\xi ' g' -\mu' g' \bigg).
\end{eqnarray}
 This two sets of equations can be solved by giving a known seed $T_{\mu \nu}$ ($\xi$ and $\mu$) and then choosing a physical condition (for instance, some of the energy conditions) to find the $\theta$-sector, $\theta_{\mu \nu}$ ($\lambda$, $f$ and $g$). In the next section we describe how to construct a hairy black hole by imposing energy conditions as supplementary constraints to the system.

\section{Hairy Black Hole}\label{HBH}
In this section we shall construct a hairy black hole by deforming the Schwarzschild metric (the seed), given by
\begin{eqnarray}\label{Schwarzschild}
e^{\xi}=e^{-\mu}=1-\frac{2M}{r},
\end{eqnarray}
with deformation functions $\{f,g\}$ chosen in a suitable way. First, let us assume that the deformed solution 
has an event horizon which we demand is given by
\begin{eqnarray}\label{horizon0}
e^{\nu(r_{H})}=e^{-\lambda(r_{H})}=0,
\end{eqnarray}
with $r_{H}$ the horizon radius. This requirement can be trivially satisfied by demanding the Schwarzschild condition
\begin{eqnarray}\label{horizon}
e^{\nu}=e^{-\lambda},
\end{eqnarray}
everywhere, which entails
\begin{eqnarray}\label{eqstate}
\Tilde{p}_r=-\Tilde{\rho}.
\end{eqnarray}
Another consequence of Eq. \eqref{horizon} is that the geometric deformation function $f$ can be written as
\begin{eqnarray}
 f=\bigg( 1-\frac{2M}{r} \bigg) ( e^{ g} -1),
\end{eqnarray}
which leads to
\begin{eqnarray}\label{metricdeformed}
ds^{2}&=&\bigg( 1-\frac{2M}{r} \bigg) h(r) dt^{2} -\bigg( 1-\frac{2M}{r} \bigg)^{-1} h^{-1}(r) dr^{2} \nonumber\\
&&-r^{2}d\Omega^{2},
\end{eqnarray}
where $h(r)=e^{ g(r)}$ is defined for future convenience.

The next step consists in to impose an extra constraint to solve for the auxiliary function $h$. In Ref. \cite{Ovalle:2020kpd}
this extra condition was to demand either the SEC or the DEC on and outside the horizon. However, it is well known that some physically relevant black holes satisfy a less restrictive condition as the WEC (see \cite{Balart:2014cga}, for example). This is the route that we shall take in this work, namely, we will demand that our solution satisfies the WEC,
\begin{eqnarray}\
&&\Tilde{\rho}\geq 0, \nonumber \\
&&\Tilde{\rho}+\Tilde{p}_r\geq 0,\label{WEC}\\
&&\Tilde{\rho}+\Tilde{p}_t\geq 0 \nonumber
\end{eqnarray}
which, after considering Eq. \eqref{eqstate}, reduces to
\begin{eqnarray}
\theta_0^0 \geq 0, \label{WEC1} \\
\theta_0^0 \geq \theta_2^2.\label{WEC2}
\end{eqnarray}
Using Eqs. \eqref{T0source} and \eqref{T2source} the inequalities can be rewritten as 
\begin{eqnarray}
1-h-(r-2M)h' \geq 0, \label{ineq1} \\
2-2h+4Mh'+r(r-2M)h'' \geq 0.\label{ineq2}
\end{eqnarray}

Note that we can find some $h$ fulfilling \eqref{ineq1} whenever
\begin{eqnarray}
1-h-(r-2M)h'=G(r),
\end{eqnarray}
for certain $G>0$. The general solution for this equation reads
\begin{eqnarray}
h(r)=\frac{r-c_1}{r-2M} - \frac{1}{r-2M} \int G(x)dx,
\end{eqnarray}
where $c_1$ is a constant with dimensions of length. There is an extra constraint on $G(r)$ as a consequence of condition \ref{ineq2} given by
\begin{eqnarray}\label{G}
2 G-r G' \geq 0.
\end{eqnarray}
Although $G$ could be an arbitrary positive function satisfying (\ref{G}), in this work we shall take 
\begin{eqnarray}
G(r)=\alpha \frac{M}{r^2} \ln \bigg( \frac{r}{\beta} \bigg)
\end{eqnarray}
from where
\begin{eqnarray}
1-h-(r-2M)h'=\alpha \frac{M}{r^2} \ln \bigg( \frac{r}{\beta} \bigg) \geq 0, \label{ODE} 
\end{eqnarray}
Where $\beta$ and $\alpha$ are constants with dimensions of a length and
\begin{eqnarray} \label{alfacondition}
\alpha \geq 0. 
\end{eqnarray}

The solution of Eq. \eqref{ODE} is
\begin{eqnarray}\label{ODEsolution}
h(r)=\frac{r-c_1}{r-2M} + \frac{\alpha M}{r(r-2M)} \bigg(1+\ln \bigg( \frac{r}{\beta}\bigg) \bigg),
\end{eqnarray}
where $c_1=2M$ in order to recover the Schwarzschild solution when $\alpha\to0$. Note that the inequality (\ref{ineq2}) is satisfied 
whenever
\begin{eqnarray}\label{betac}
\ln\bigg( \frac{r}{\beta}\bigg)\geq \frac{1}{4}.
\end{eqnarray}
Now, replacing Eq. \eqref{ODEsolution} in \ref{metricdeformed} we arrive at
\begin{eqnarray}\label{metricaT}
e^\nu=e^{-\lambda}=1-\frac{2M}{r}+\frac{\alpha M}{r^2}+\frac{\alpha M}{r^2} \ln \bigg( \frac{r}{\beta} \bigg).
\end{eqnarray}
At this point some comments are in order. First,
note that the solution (\ref{metricaT}) coincides with the Reissner-Nordstr\"{o}m (RN) solution if the logarithmic term is absent and after defining the electric charge as $Q\equiv \sqrt{\alpha M}$. Indeed, we could interpret our solution as a result of certain non--linear electrodynamics source. However, as the presence of the logarithmic term 
makes the solution not to be asymptotically RN, we prefer to consider
such an identification as merely formal. In this regard, we could take the parameter $\alpha$ as genuine hair which is not associated with any global charge, namely the mass, the (Maxwellian) electric charge or the angular momentum. In this sense, the logarithmic term is of fundamental importance, which could be used to justify our choice for $G(r)$. What is more, there is an screening effect due to the massive term in the sense that, for an asymptotic observer the metric is given by
\begin{eqnarray}
e^{\nu}=1-\frac{2M}{r}+\mathcal{O}(r^{-2}),
\end{eqnarray}
so that the solution is indistinguishable from the Schwarzschild black hole with the same mass. This occurs for example in the Einstein--Yang--Mills  and Skyrmions models \cite{Bizon:1994dh}. Second, we can accommodate the free parameter $\beta$ to ensure that our hairy solution has the same horizon as the Schwarzschild BH, namely $r_{H(S)}=2M$. Doing so, we obtain $\beta=2Me$ with $e$ the Euler's number. However, the condition (\ref{betac}) implies that  the solution satisfies the WEC outside the horizon, specifically at $r>e^{5/4}r_{H(S)}$. As we are looking for a hairy solution satisfying the WEC at least at and outside the horizon,
we can take advantage of the formal similitude of our solution with the RN BH and take 
\begin{eqnarray}\label{beta}
\beta =r_H e^{-\xi},
\end{eqnarray}
with $r_{H}$ the horizon radius and $\xi\ge 1/4$, necessarily. Although we can take any value of $\xi\ge 1/4$, in this work we shall take $\xi=1/4$ for which the WEC saturates at the horizon and $r_{H}$ is given by
\begin{eqnarray}\label{metricsolution}
r_H=M \pm \sqrt{M^2- \frac{5}{4}\alpha M},
\end{eqnarray}
where $0<\alpha<\frac{4}{5}M$ necessarily. With this choice, our hairy BH is characterized by the metric
\begin{eqnarray}
e^\nu=e^{-\lambda}=1-\frac{2M}{r}+\frac{5}{4}\frac{\alpha M}{r^2}+\frac{\alpha M}{r^2} \ln \bigg( \frac{r}{r_H} \bigg),
\end{eqnarray}
with a generic matter sector given by
\begin{eqnarray}\
\Tilde{\rho}&=&\theta_0^0= \frac{\alpha M}{r^4} \bigg( \frac{1}{4} +\ln \bigg( \frac{r}{r_H}\bigg) \bigg), \\
\Tilde{p}_r &=&-\theta_1^1=-\frac{\alpha M}{r^4} \bigg( \frac{1}{4} +\ln \bigg( \frac{r}{r_H}\bigg) \bigg),\\
\Tilde{p}_t &=& -\theta_2^2 =-\frac{\alpha M}{4r^4} \bigg( 1-4\ln \bigg( \frac{r}{r_H}\bigg) \bigg).
\end{eqnarray}

Finally, it is noticeable that the solution satisfies also both the SEC and the DEC for $r\ge r_{H}$.

\section{Periastron advances and gravitational lensing}\label{GM}
In this section we will explore how the hair affects the geodesic motion of 
test particles in comparison with the Schwarzschild case. To this end, we will study both the periastron advances of bounded orbits and the radius of the photon sphere.

\begin{figure}[htb!]
\centering
\includegraphics[width=0.5\textwidth]{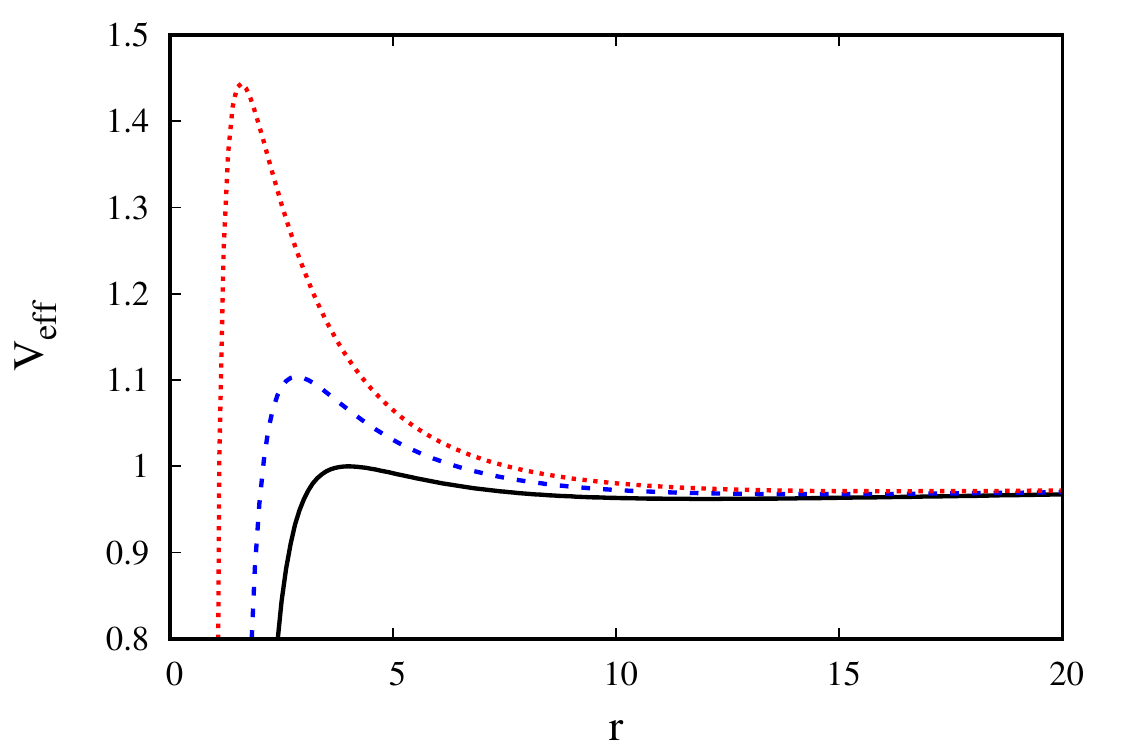}\
\caption{ 
 Effective potential $V_{eff}$ with $L=4$ for $M=1$, and $\alpha=0$ (black line, Schwarzschild case), $\alpha=0.5$ (blue dashed line) and $\alpha=1$ (red dotted line). \label{LE} }
\end{figure}

We will proceed as usual by solving the geodesic equations  
\begin{eqnarray} 
\Dot{r}^2&=&E^2-\frac{F}{r^{2}}\bigg( \mathcal{Q}+L^2 +\epsilon r^{2} \bigg) \label{rgeo}\\
\dot{\theta}&=&\frac{1}{r^{4}}(\mathcal{Q}-L^{2}\cot^{2}\theta)\\
\dot{\phi}&=&\frac{L}{r^{2}}\csc^{2}\theta\label{figeo}\\
\dot{t}&=&\frac{E}{F},
\end{eqnarray}
in the equatorial plane ($\theta=\pi/2$ and $\dot{\theta}=0$). In the equations above,
$E$ is the energy, $L$ the angular momentum, $\epsilon=0$ for null geodesics and $\epsilon=1$ for timelike geodesics, $\mathcal{Q}$ is the Carter constant and $F$ is the metric function which, in our case, is given by
\begin{eqnarray}\label{F}
F=1-\frac{2M}{r}+\frac{5}{4}\frac{\alpha M}{r^2}+\frac{\alpha M}{r^2} \ln \bigg( \frac{r}{r_H} \bigg)
\end{eqnarray}
From (\ref{rgeo}) we identify the effective potential
\begin{eqnarray} \label{Veff}
V^2_{eff}=\frac{F}{r^2}\bigg( \epsilon + \frac{L^2}{r^2} \bigg),
\end{eqnarray}
which is shown in Fig. \ref{LE} for different values of the parameters involved. Note that maximum of the potential increases as $\alpha$ grows. It is worth noticing that this behaviour differs with that found in a previous work \cite{Ramos:2021jta} where it was analysed the effective potential for the four models satisfying the SEC and the DEC reported in \cite{Ovalle:2020kpd}. Indeed, in that case, the maximum of the potential decreases with the value of the hairs. Also note that the location of the maximum shift to lower values of the radius in comparison to the Schwarzschild case.

\begin{figure*}[hbt!]
\centering
\includegraphics[width=0.3\textwidth]{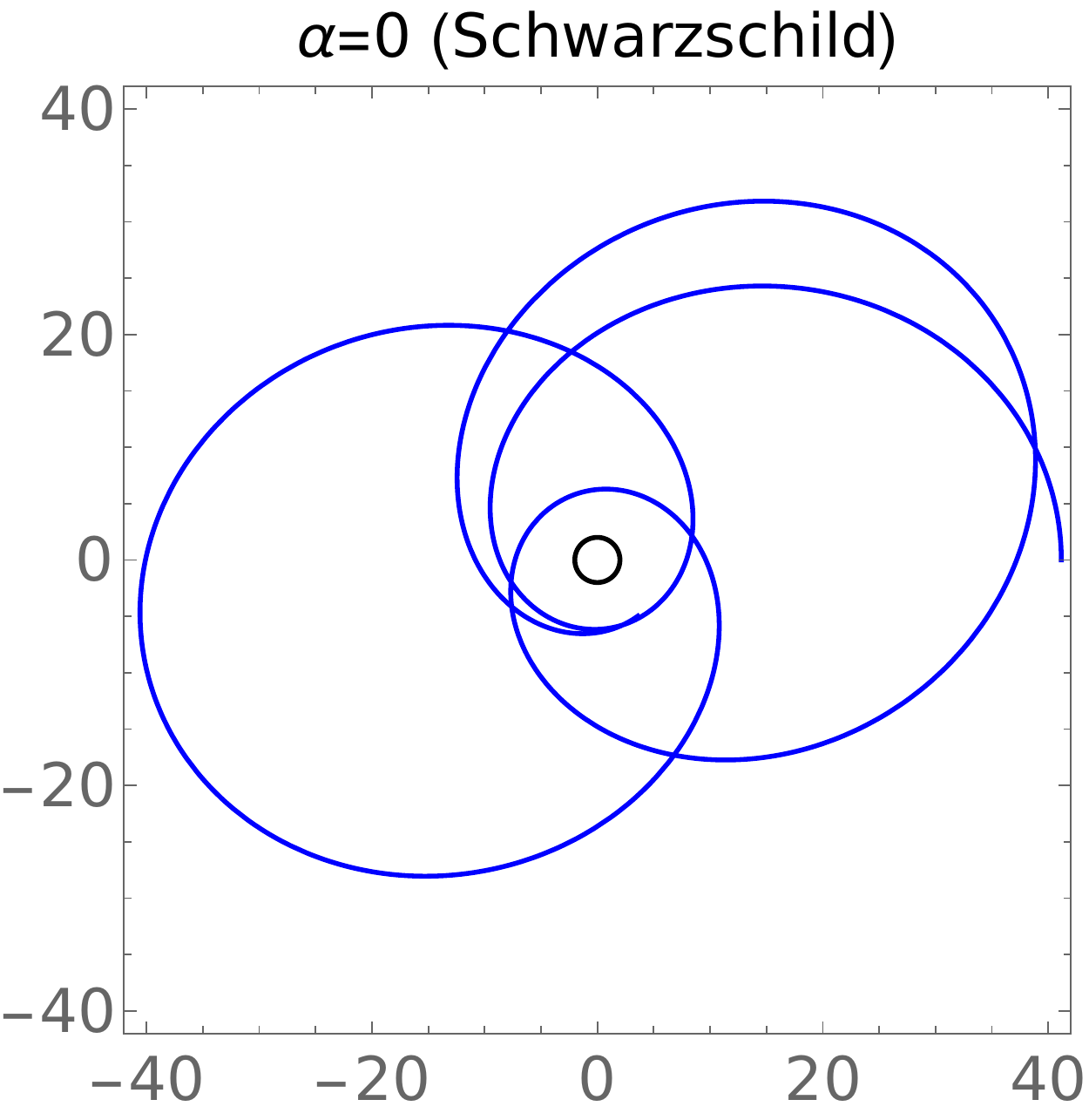}  \
\includegraphics[width=0.3\textwidth]{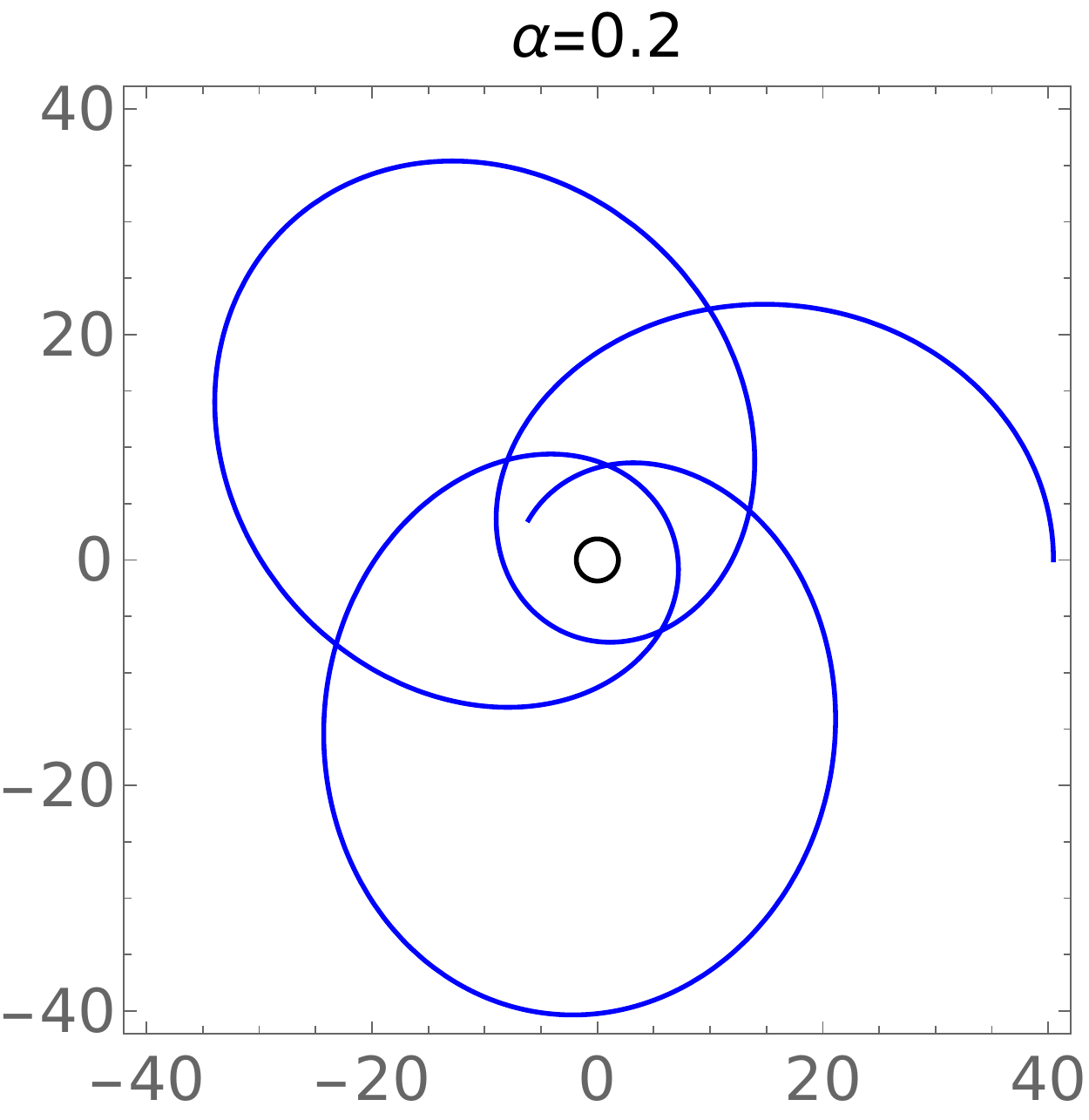}  \
\includegraphics[width=0.3\textwidth]{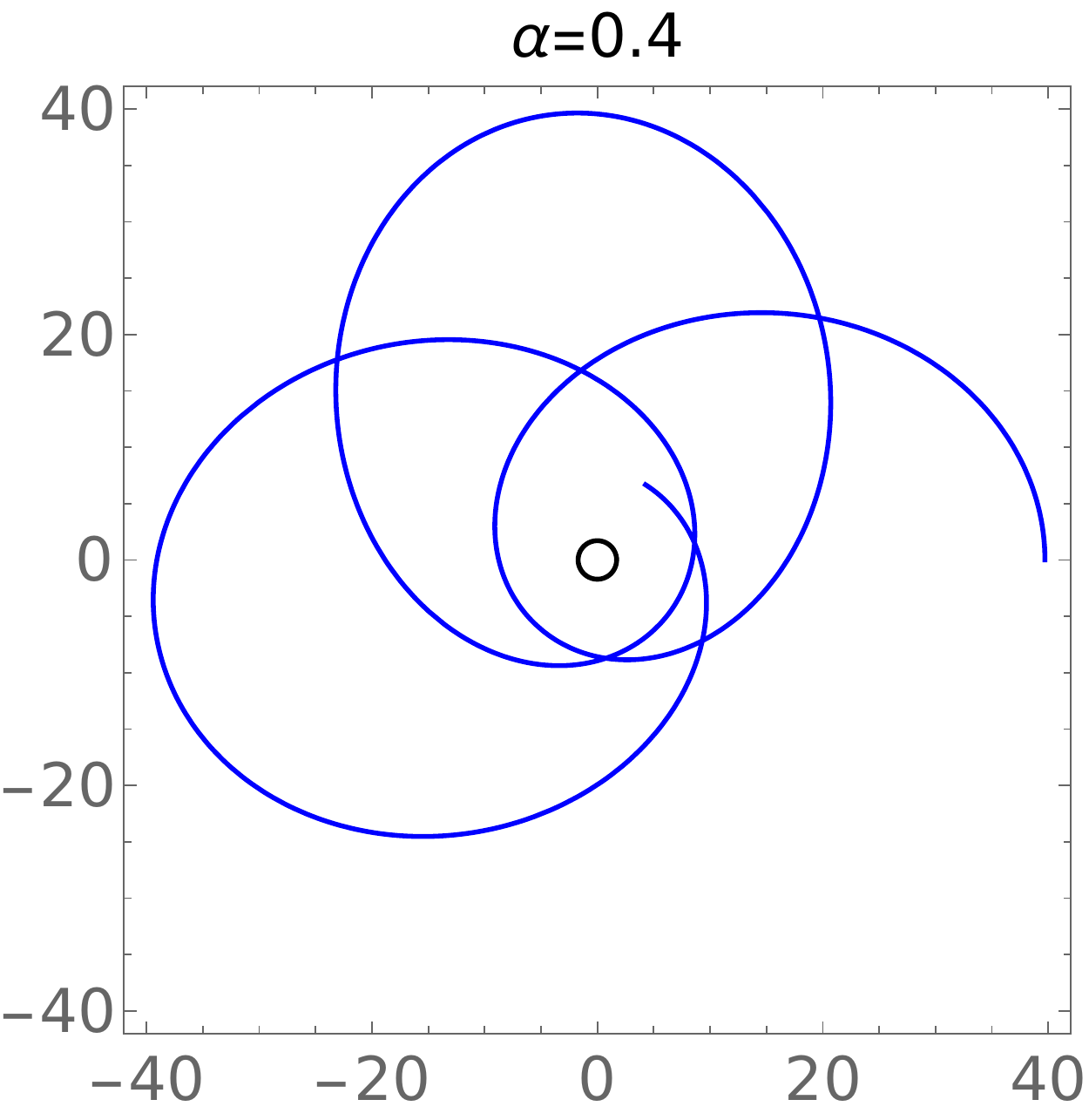}  \
\medskip
\includegraphics[width=0.3\textwidth]{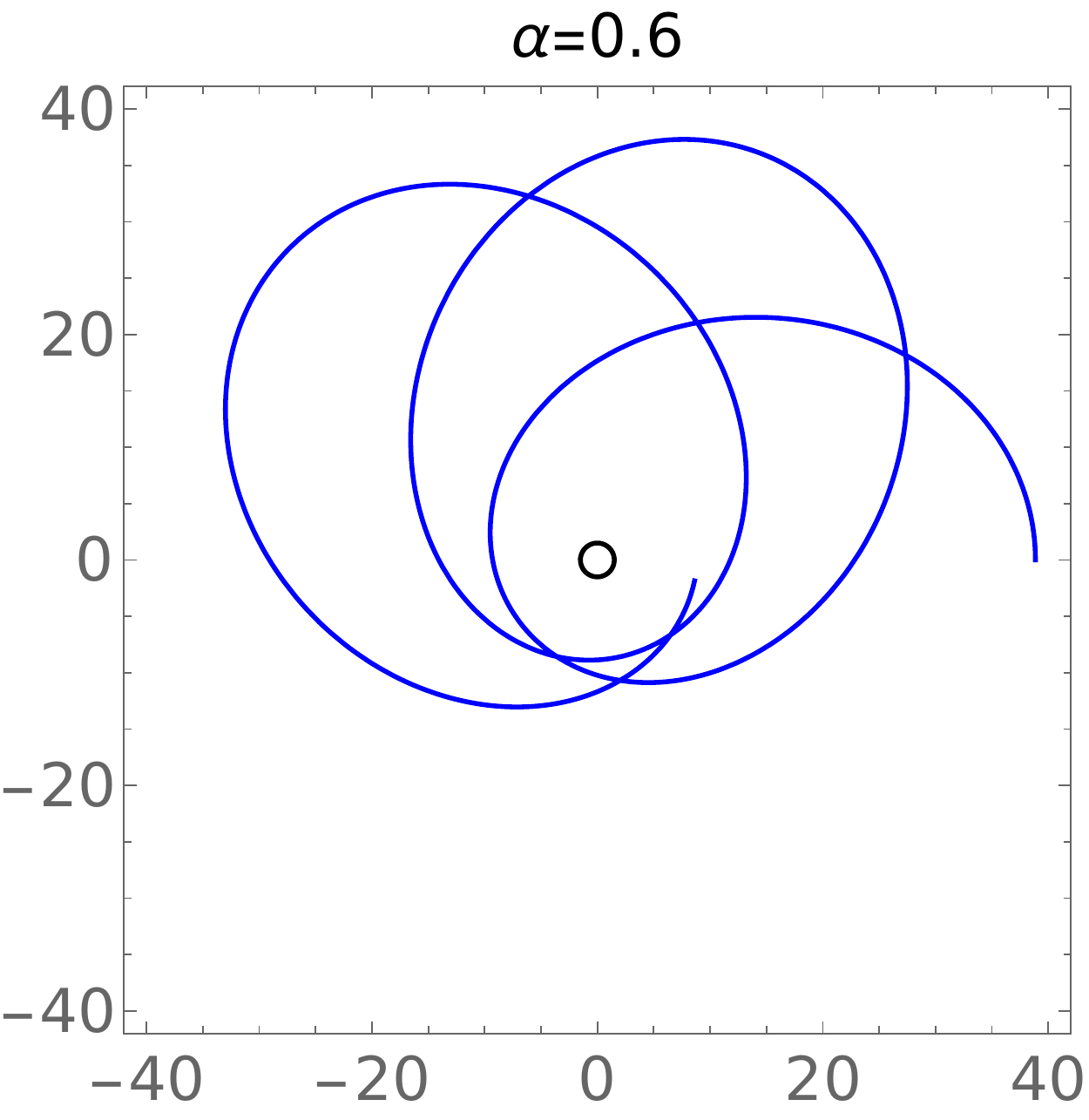}  \
\includegraphics[width=0.3\textwidth]{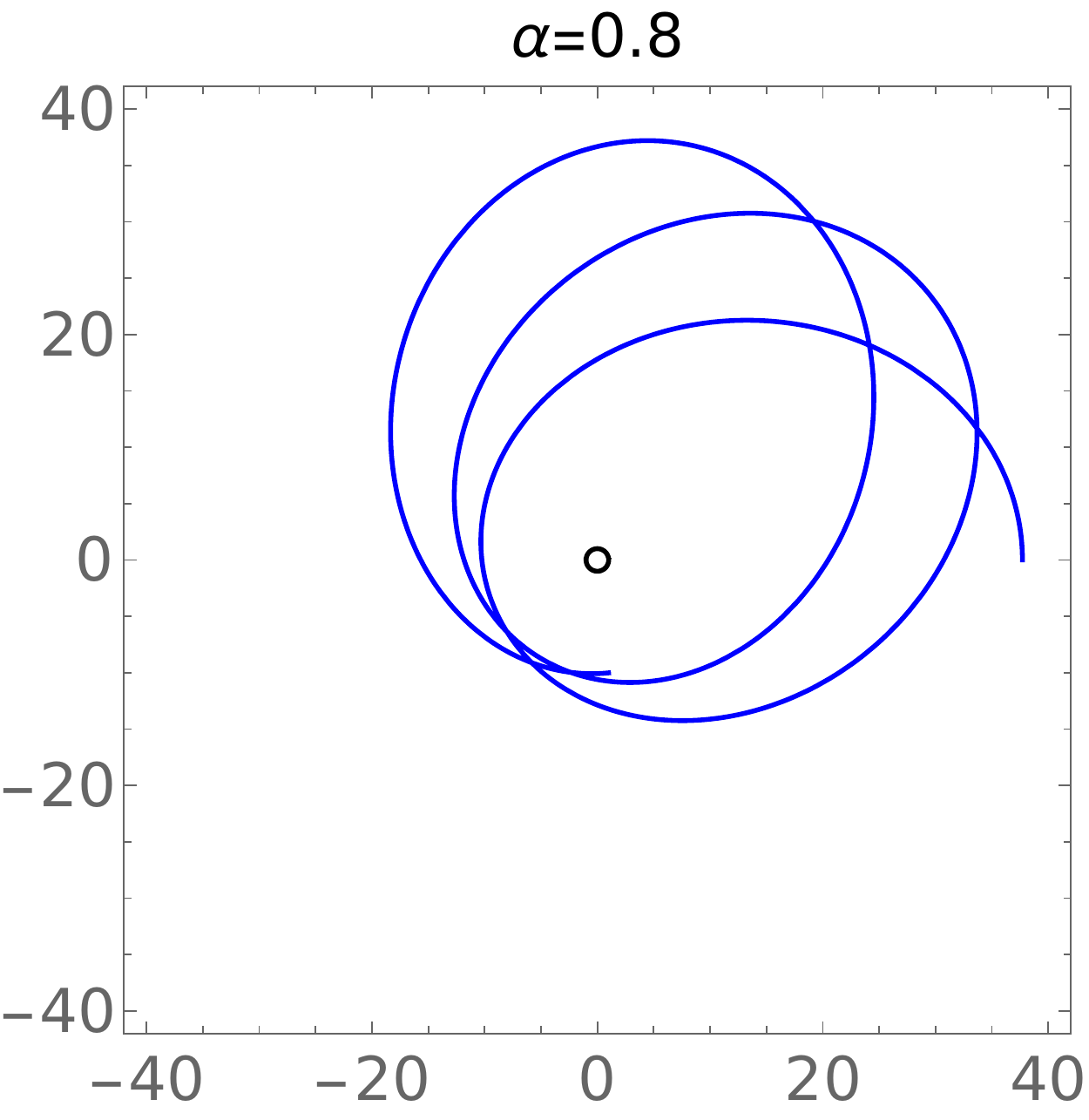}  \
\caption{Periastron advance for  $L=4$, $M=1$, and $E=0.98$ . The angle of the advance decreases as the value of the hair  grows.
\label{BO} }
\end{figure*}

For the study of the periastron advance we have to explore the behaviour of bounded orbits of massive particles ($\epsilon=1$); namely, those with two turning points obtained from the condition $E=V_{eff}$ at the convex sector. More precisely, in this work we solved Eqs. (\ref{rgeo}) and (\ref{figeo}) numerically, for
$L=4$ and $E\sim 0.9$. The behaviour of $r(\phi)$ is shown Fig. \ref{BO} where we note that the effect of the hair is to reduce the advance of the periastron. At this point we can make a qualitative comparison with the models in \cite{Ramos:2021jta} in terms of the size of the shadow. Note that, for approximately the same values of the parameters, the size of the shadow for the model here is bigger that the found in \cite{Ramos:2021jta} for all the metrics under consideration.

\begin{figure}[htb!]
\centering
\includegraphics[width=0.5\textwidth]{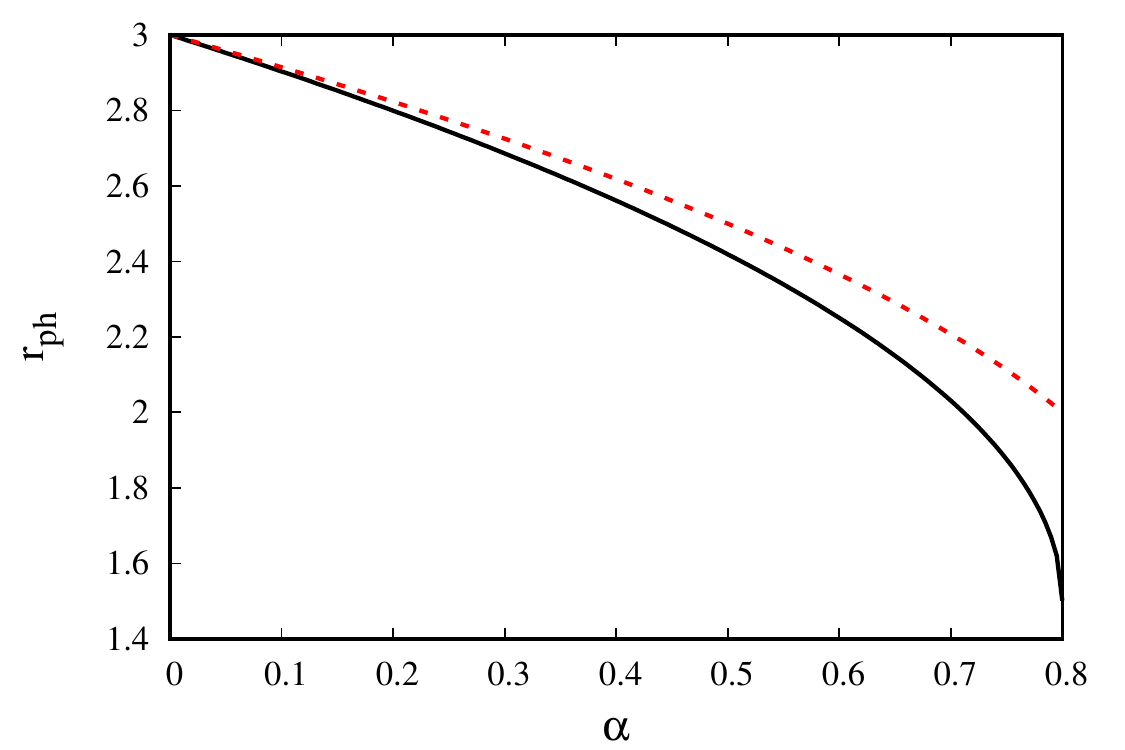}\
\caption{Radius of the photon sphere, $r_{ph}$, as a function of $\alpha$ for $M=1$. The red dashed line corresponds to RN case. Note that $r_{ph}$ decreases faster than $r_{ph}^{RN}$ \label{photon}} 
\end{figure}

The effect of the hair on the gravitational lensing can be analysed in terms of the radius of the photon sphere that is obtained numerically from Eq. (\ref{rgeo}) with $\epsilon=0$ by imposing $\dot{r}=0$ which leads to
\begin{eqnarray}
r=\sqrt{F}\frac{L}{E}.
\end{eqnarray}
In Fig. \ref{photon}, it is shown the radius of the photon sphere $r_{ph}$ as a function of $\alpha$ for $M=1$ (black line). Note that $r_{ph}$ decreases as $\alpha$ grows which entails a diminishing in the size of the black hole shadow as a consequence of the hair. Furthermore, this behaviour is similar to what occurs in the RN case where the radius of the photon sphere is given by
\begin{eqnarray}
r_{ph}^{RN}=\frac{3M}{2}\left(1+\sqrt{1-\frac{8}{9}\left(\frac{Q}{M}\right)^{2}}\right),
\end{eqnarray}
which is always smaller than the Schwarzschild radius for any $Q\ne0$. However, the decreasing of $r_{ph}$ is faster in our case in comparison with the RN BH as shown in Fig. \ref{photon}
Another interesting point which deserves to be mentioned, is that behaviour if $r_{ph}$ as shown in Fig. \ref{photon} coincides qualitatively with the model 2 in Ref. \cite{Ramos:2021jta} which corresponds to a hairy black hole satisfying the WEC. For the other models, $r_{ph}$ is an increasing function of the hair parameters.

\section{Impact Parameters and Orbits}\label{IP}
\subsection{Impact Parameters}
\begin{figure*}[hbt!]
\centering
\includegraphics[width=0.45\textwidth]{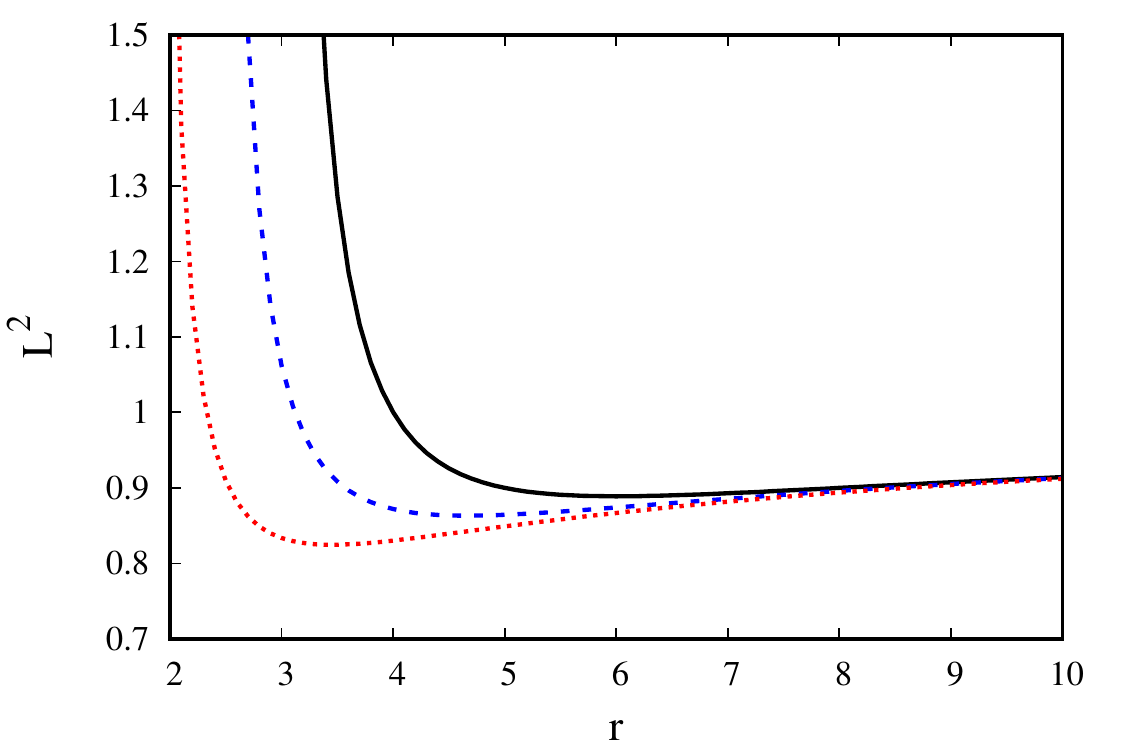}  \
\includegraphics[width=0.45\textwidth]{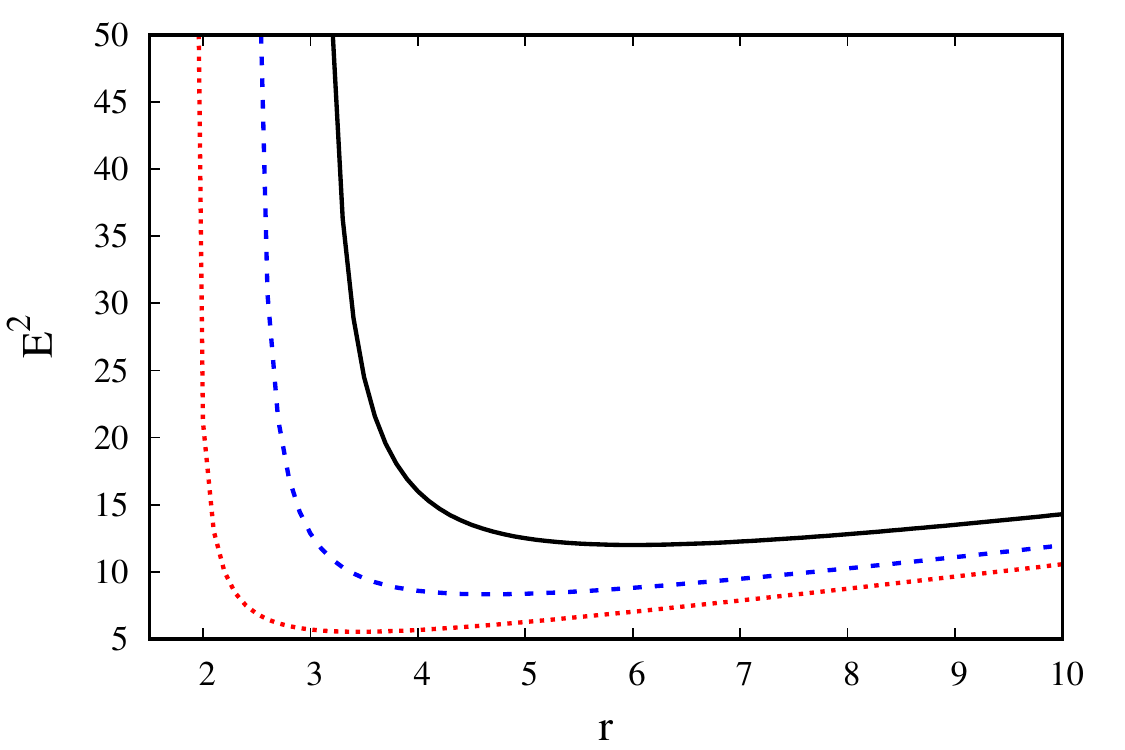}  \
\caption{Impact parameters ($L^2$, $E^2$) for $M = 1$, and $\alpha = 0$ (black line, Schwarzschild case), $\alpha = 0.5$ (blue dashed line), and $\alpha = 0.8$ (red dotted line).
\label{IPG} }
\end{figure*}
For bounded orbits we can study the energy and angular momentum of the system by imposing $\Dot{r}=0$ and $\Ddot{r}=0$ in Eq. \ref{rgeo} from where
\begin{eqnarray}
    L^2 = \frac{2 F^2}{2F-r F'},
\end{eqnarray}
\begin{eqnarray}
    E^2=\frac{2 r^2 F}{2F -r F'}-r^2,
\end{eqnarray}
with $F$ corresponding to the metric. The impact parameters are illustrated in Fig. \ref{IPG}. The location of the minimum value of $L^2$ and $E^2$ decreases as the parameter $\alpha$ increases in contrast with what occurs in Ref. \cite{Ramos:2021jta}. About $L^2$, the three curves intersect around $r\sim 15.2$ and there is an interchange of the roles being the black line the lower bound (this interval is not shown in the plot but was numerically calculated).
\subsection{ISCO}
The Innermost Stable Circular Orbit (ISCO) corresponds to the last stable orbit that occurs when tha maximum of the effective potential becomes an inflection point. For a fixed value of the mass $M$, there exits a value of angular momentum $L$ and a radius $r_{ISCO}$ which correspond to this circular orbit. To find the ISCO there is a system of two equations that must be solved numerically to find the inflection point of the effective potential, namely $V'_{eff}=0$ and $V''_{eff}=0$.
\begin{figure}[h!]
\centering
\includegraphics[width=0.5\textwidth]{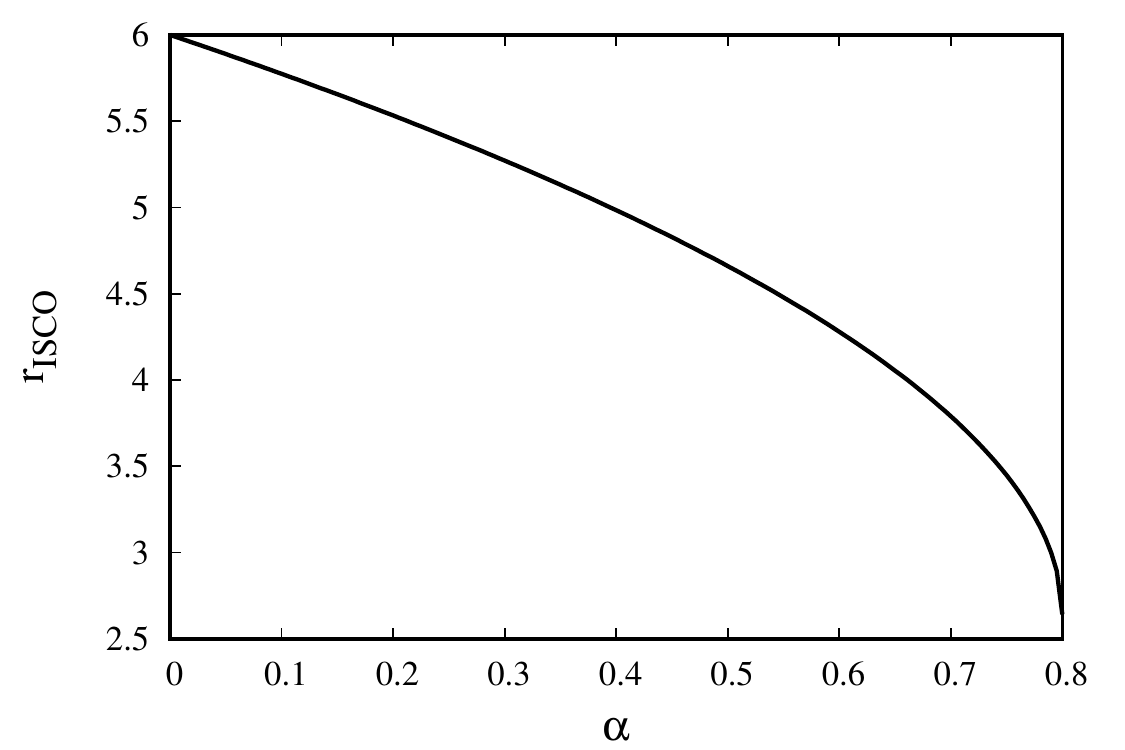}\
\caption{Innermost Stable Circular Orbit (ISCO), $r_{ISCO}$, as a function of $\alpha$ for $M=1$. \label{ISCO}} 
\end{figure}
In Fig. \ref{ISCO} is shown that the ISCO radius is as monotonously decreasing function of the hair parameter $\alpha$. Note that, for $\alpha=0$ the value of $r_{ISCO}=6$, namely the Schwarzschild case is recovered. The minimum value is $r_{ISCO}=2.64$ and is accomplished when the free parameter takes it maximum value $\alpha=0.8$. It is worth mentioning that this behaviour coincides formally with the ISCO of model 3 in Ref. \cite{Ramos}. Indeed, for model 1 $r_{ISCO}$ is an increasing function of the hair parameter and for models 2 and 4, $r_{ISCO}$ reach a minimum.
\subsection{MBO}
The Marginally Bound Orbit (MBO) corresponds to the critical bound orbit with energy $E=1$ that separate the bounded orbits ($E<1$) from the unbounded orbits ($E>1$). For a fixed value of the mass $M$, there exits a value of angular momentum $L$ and a radius $r_{MBO}$ which correspond to this last bounded orbit. In order to find $r_{MBO}$, the system $V_{eff}=1$ and $V'_{eff}=0$ must be solved numerically.
\begin{figure}[h!]
\centering
\includegraphics[width=0.5\textwidth]{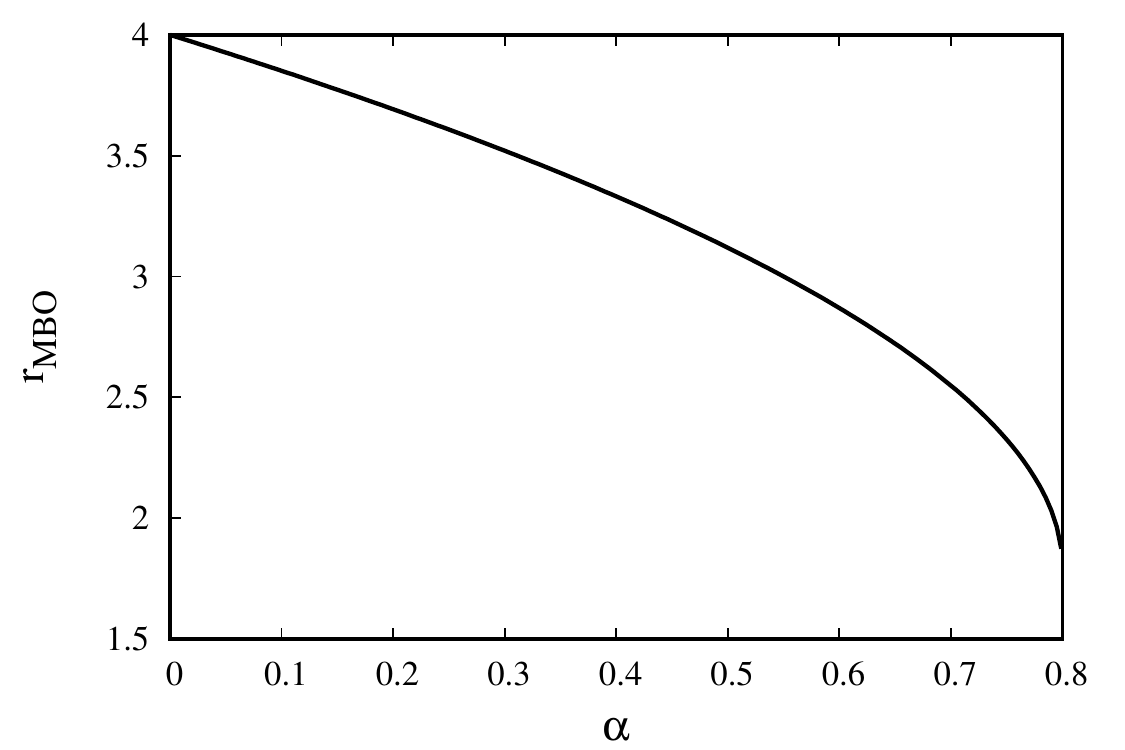}\
\caption{Marginally Bound Orbit (MBO), $r_{MBO}$, as a function of $\alpha$ for $M=1$. \label{MBO}} 
\end{figure}\\
In Fig. \ref{MBO} is shown that $r_MBO$ exhibits a concave and decreasing behaviour with. Note that, $\alpha=0$ entails $r_{MBO}=4$  (Schwarzschild case). It is worth mentioning that this behaviour coincides with the model 2 in \cite{Ramos:2021jta}. Indeed, for model 1 $r_{MBO}$ reach a minimum and for models 3 and 4 is an increasing function of the hair parameter.

\section{Quasi Normal Modes}\label{QNM}
As it is well known, the study of the evolution of test fields around a fixed BH background shed lights about the
stability of the black hole. Such {\it perturbations}, can be either scalar or vector fields or even small deviations in the space--time background (tensorial perturbations). Remarkably, whatever the path we follow, the behaviour of the perturbation field is govern by a  Schr\"{o}dinger--like equation, namely
\begin{eqnarray} \label{schrodinger}
\bigg( \frac{d^2}{dr_*^2} +\omega ^2 -V(r_*) \bigg)\chi(r_*)=0,
\end{eqnarray}
where $r_*$ is the tortoise radial coordinates
\begin{eqnarray}
\frac{dr_*}{dr}=\frac{1}{F(r)},
\end{eqnarray}
$\omega$ represents the frequency of the QNM and has a real and an imaginary part, namely $\omega=Re(\omega)+iIm(\omega)$, and $V(r)$ is an effective potential, which for axial perturbations takes the form
\begin{eqnarray} \label{veff}
V_L (r) = F(r) \bigg( \frac{L (L+1)}{r^2} +F'(r)\frac{(1-s^2)}{r} \bigg),
\end{eqnarray}
where $s=0,1,2$ is the spin of the perturbing field and $F$ is the metric function given by (\ref{F}). One interesting feature about QNM's is that they dominate in the decay of perturbations at late times and, therefore, are observed by gravitational interferometers \cite{Abbott:2016blz,Abbott:2017oio,Abbott:2017gyy,Konoplya:2022hll}. Besides, they can be connected to Event Horizon Telescope data \cite{Akiyama:2019cqa,Akiyama:2019bqs} given the relationship between the real part of the QNM frequencies and the shadow cast by the BH as demonstrated in \cite{Cuadros-Melgar:2020kqn} at $3^{rd}$  order WKB. Indeed, at the eikonal regime, the radius of the shadow, $\mathcal{R}_{s}$ and $Re(\omega)$ reads \cite{Cuadros-Melgar:2020kqn}
\begin{eqnarray}
Re(\omega)=\mathcal{R}_{s}^{-1}\left(\ell+\frac{1}{2}\right).
\end{eqnarray}

In this work, we shall study the evolution of the perturbation field for each value of $s$, namely, we shall consider scalar ($s=0$), vectorial ($s=1$) and tensorial ($s=2$) perturbations. The solutions of Eq. (\ref{schrodinger}) with the appropriate boundary conditions, namely
\begin{eqnarray} \label{bc}
\chi (r_*) \sim C_{\pm}\exp (\pm i \omega r_*), \,\,\,\,\, r_*\rightarrow \pm \infty,
\end{eqnarray}
(which corresponds to demand purely out--going waves when approach to infinity) carries information of the QNM frequencies $\omega$. In the last years, several strategies have been implemented to solve (\ref{schrodinger}) (for an incomplete list see \cite{Konoplya:2022tvv,Churilova:2021nnc,Konoplya:2020jgt,Konoplya:2020bxa,Konoplya:2019nzp,Rincon:2021gwd,Panotopoulos:2020mii,Rincon:2020cos,Rincon:2020iwy,Rincon:2020pne,Xiong:2021cth,Zhang:2021bdr,Panotopoulos:2019gtn,Lee:2020iau,Churilova:2019qph,Oliveira:2018oha,Blazquez-Salcedo:2018ipc,Panotopoulos:2017hns} and references therein, for example). Nevertheless, in this work we shall implement the semi--analytical WKB approximation reported in \cite{Konoplya:2019hlu} inspired in the study scattering around BH's, given its similarity with the one--dimensional Schr\"{o}dinger equation with a potential barrier reported in \cite{Schutz:1985km}. To be more specific, in \cite{Konoplya:2019hlu} is obtained that the $13^{th}$ WKB order formula is given by
\begin{eqnarray} \label{WKB}
i \frac{\omega ^2 -V_0}{\sqrt{-2V_0''}}-\sum_{j=2}^{k}\Lambda_j=k+\frac{1}{2},
\end{eqnarray}
where, $k$ is the order of the WKB, $V_0$ is the maximum height of the potential and $V_0''$ is its second derivative with respect to the tortoise coordinate evaluated at the radius where $V_{0}$ reaches a maximum. The values $\Lambda_j$ are corrections that depend on the value of the potential and higher derivatives of it at the maximum. The exact expressions for the terms $\Lambda_j$ are too long to be shown here but can be found in  \cite{Konoplya:2019hlu}. It should be emphasized that, an increasing in $k$ does not necessarily lead to a better approximation of the quasinormal frequencies. To be more precise, the order $k$ leading to the best value of $\omega$ is not unique but could depend on the pair $(n,L)$ chosen. For that reason, in this work we shall implement the following strategy for the computation of the best $\omega$ for each order:\\ 
1. We estimate the accuracy by using \cite{Konoplya:2019hlu}
\begin{equation} \label{Deltak}
\Delta_k= \frac{| \omega_{k+1}-\omega_{k-1} |}{2},
\end{equation}
which, as discussed in \cite{Konoplya:2019hlu}, is usually greater than the error, namely
\begin{eqnarray}\label{errorD}
\Delta_{k}\geq |\omega-\omega_{k}|.
\end{eqnarray}
with $\omega$ the accurate value of the quasinormal frequency. From (\ref{errorD}) we see that as $\Delta_{k}\to0$ for some $k$, the accuracy of the WKB increases. \\
2. We compute $Im(\omega)$ with the smallest $\Delta_{k}$ associated to a given pair $(n,L)$.\\
\\

\begin{figure*}[hbt!]
\centering
\includegraphics[width=0.3\textwidth]{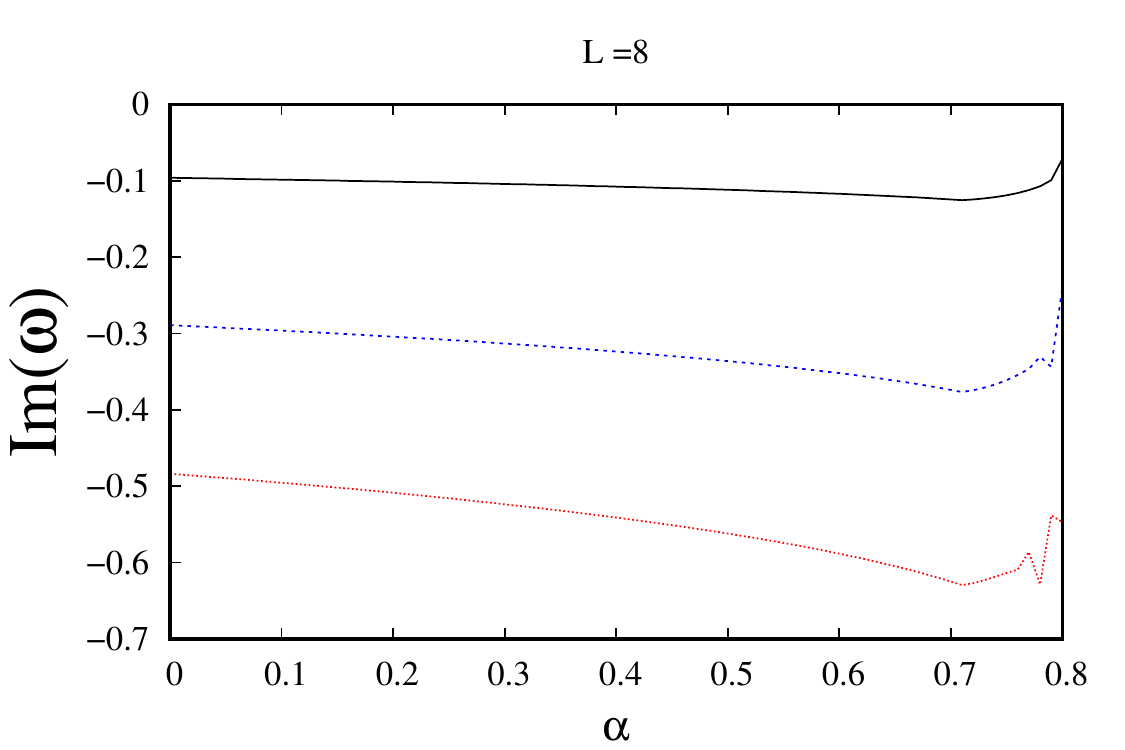}  \
\includegraphics[width=0.3\textwidth]{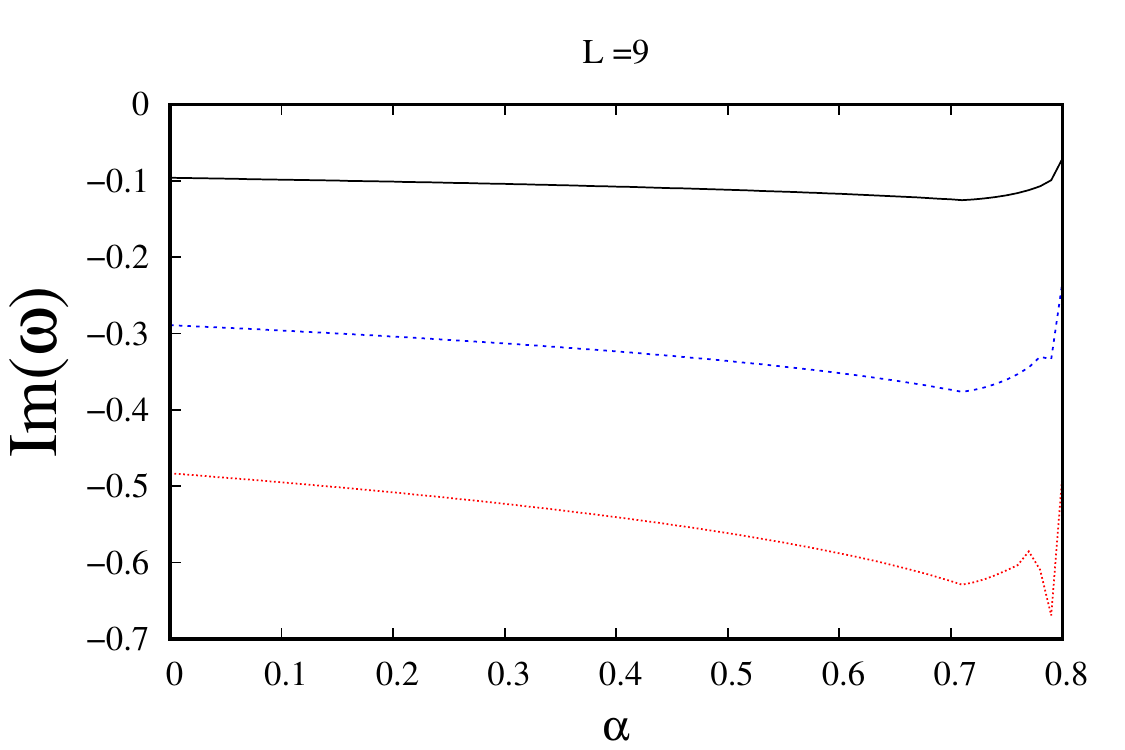}  \
\includegraphics[width=0.3\textwidth]{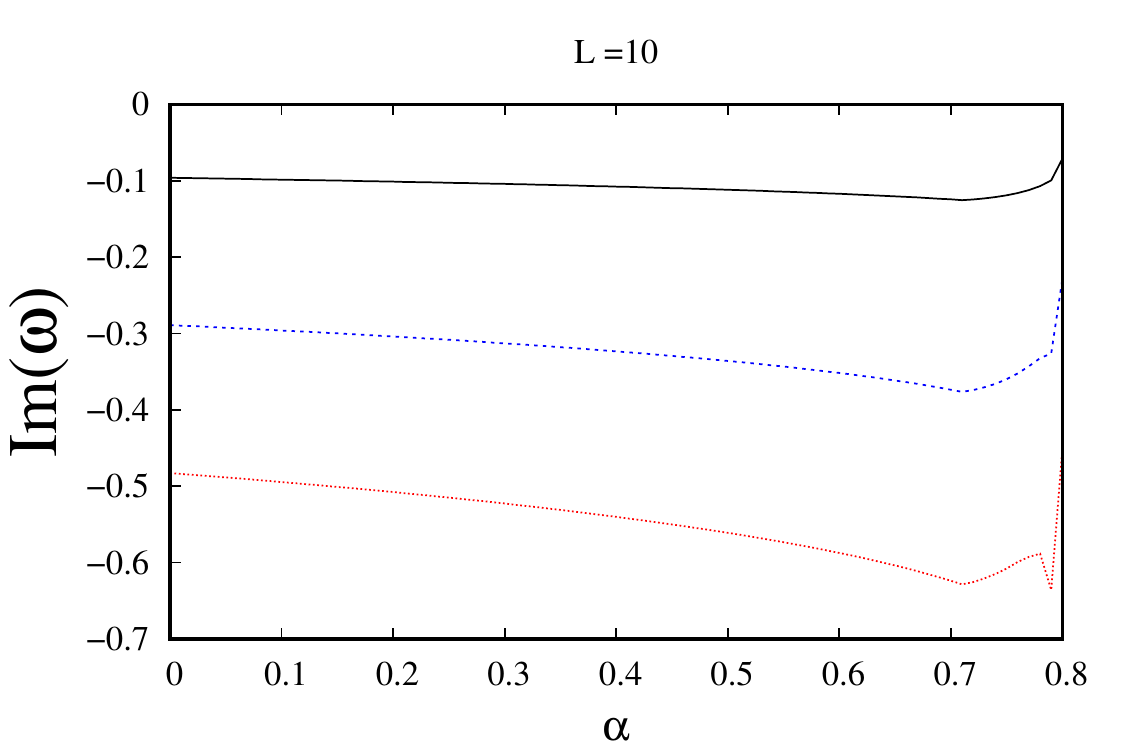}  \
\medskip
\includegraphics[width=0.3\textwidth]{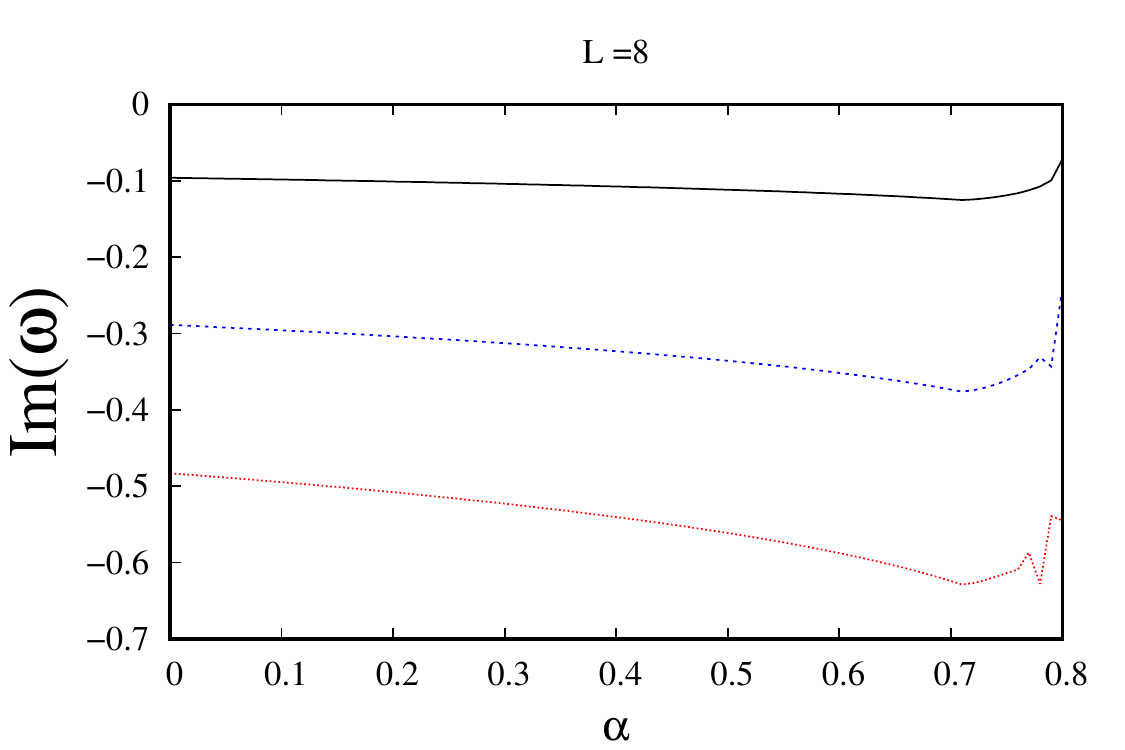}  \
\includegraphics[width=0.3\textwidth]{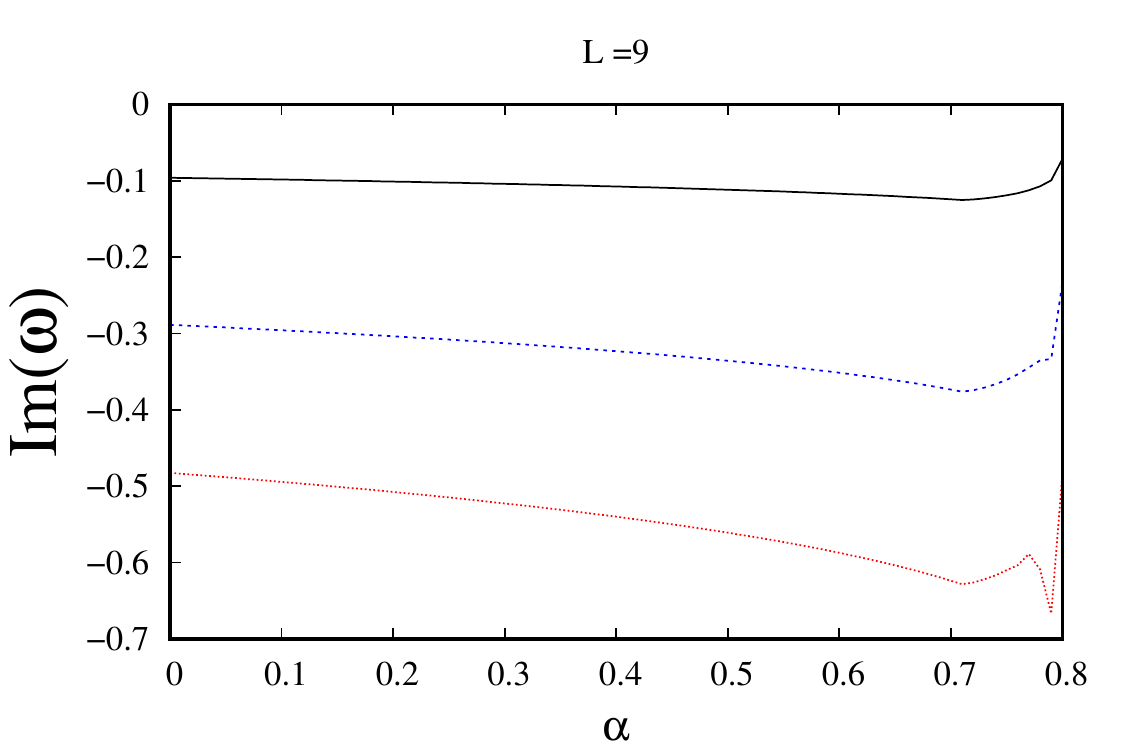}  \
\includegraphics[width=0.3\textwidth]{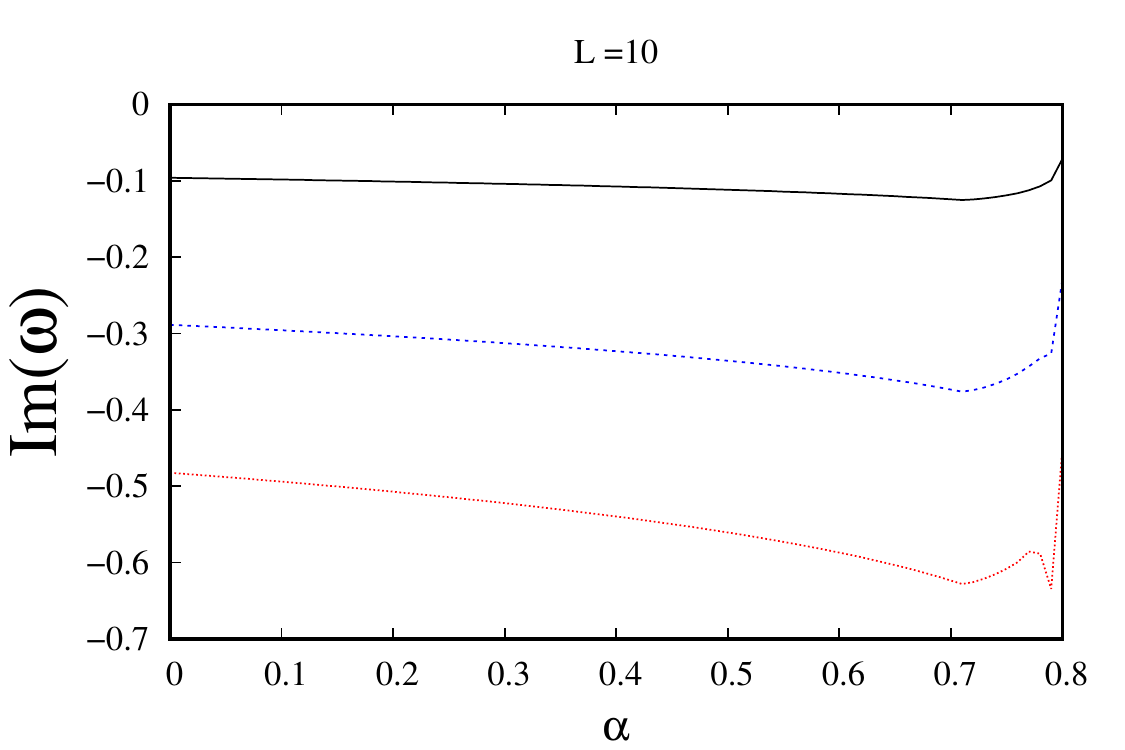}  \
\medskip
\includegraphics[width=0.3\textwidth]{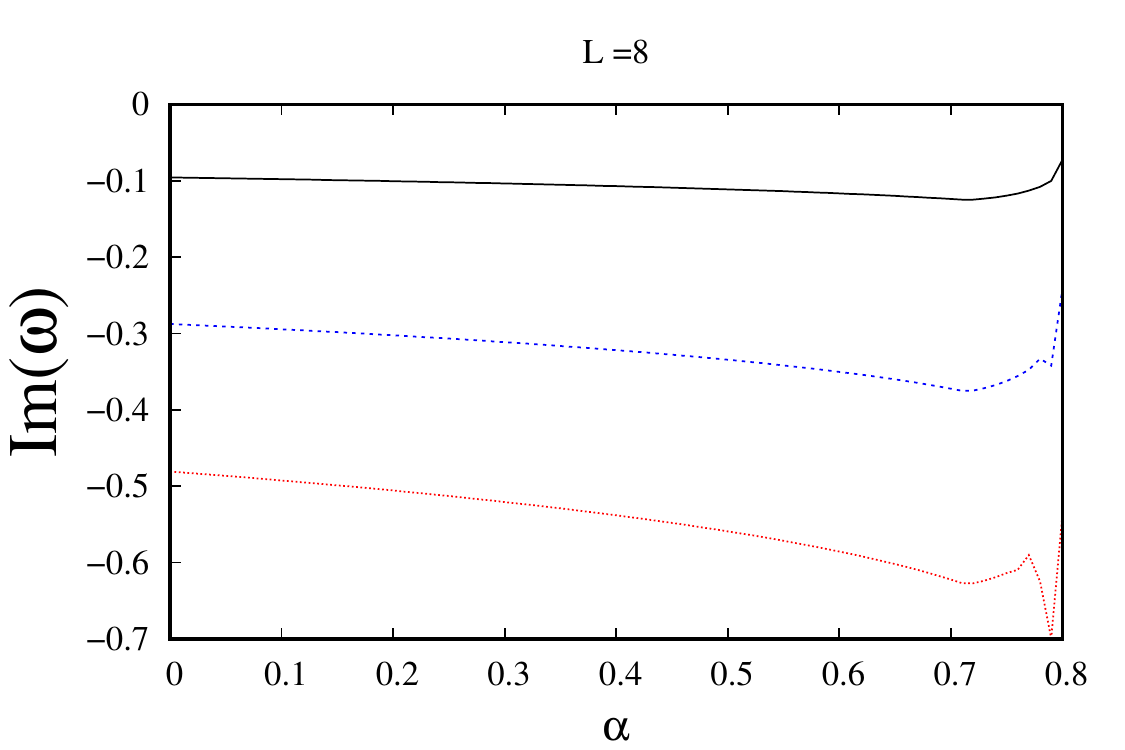}  \
\includegraphics[width=0.3\textwidth]{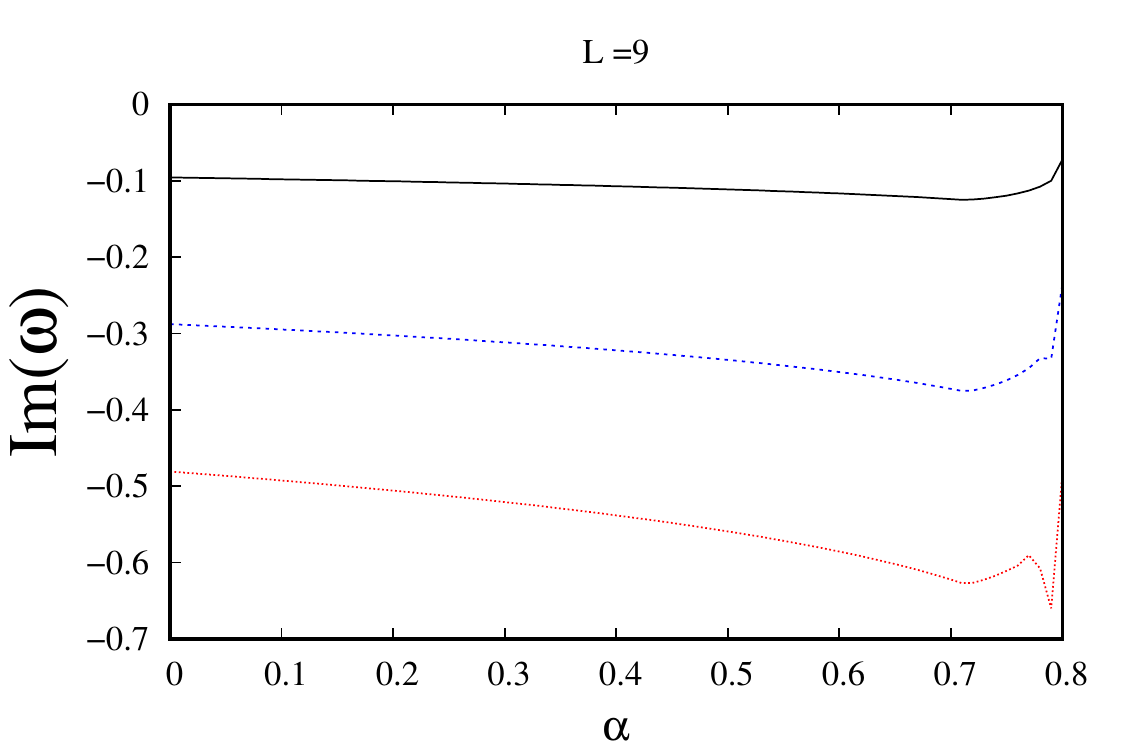}  \
\includegraphics[width=0.3\textwidth]{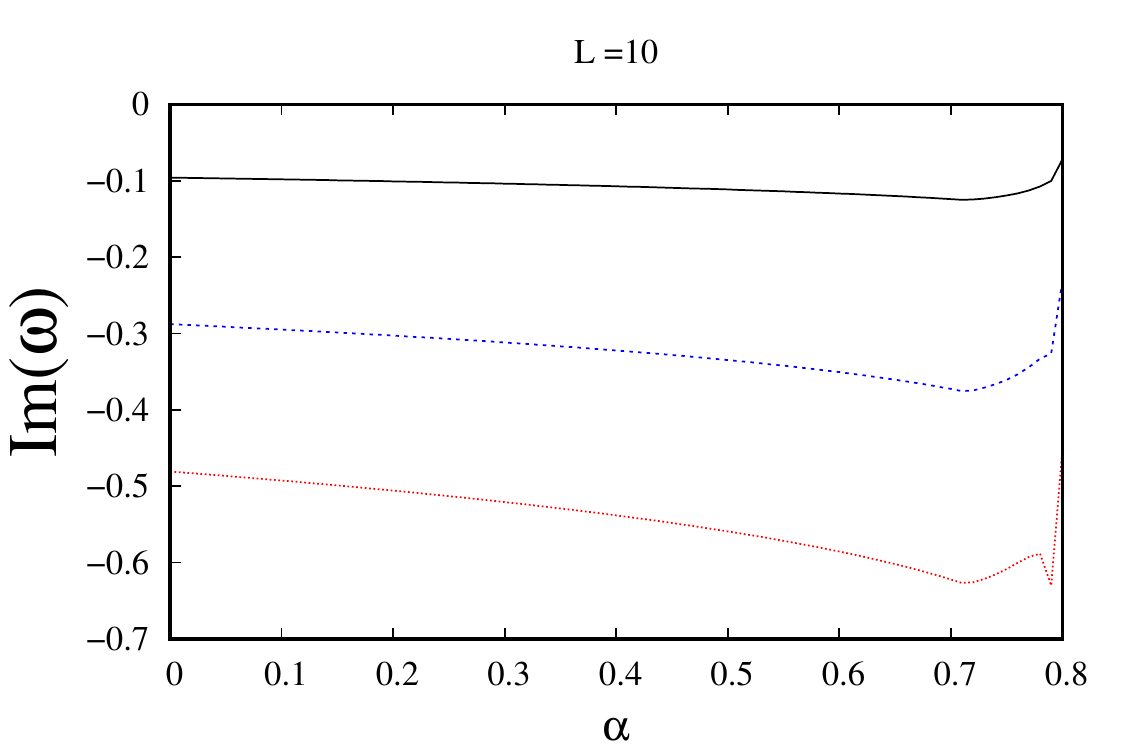}  \
\caption{\label{imw} $Im(\omega)$ as a function of $\alpha$ for $\ell=8,9,10$ and $n=0$ (black line), $n=1$ (blue line) and $n=2$ (red line) for $s=0$ (frist row), $s=1$ (second row) and $s=2$ (third row).}
\end{figure*}

In Table \ref{table:1} we show $\Delta_{k}$ for each of the perturbing fields under consideration, namely $s=0,1,2$ for $\ell=8,9,10$, $n=0,1,2$, and $k=(2,3...,7)$. We have also compute (not shown here) $\Delta_{k}$ for smaller $\ell$'s ($\ell=0,1,2,..,7$) and greater $k$'s ($k=8,9,..12$) but as the error increases considerably, we consider this data as not confident. In any case, in Table \ref{table:1} it is shown that, in this case, the $6^{th}$ order is the one with the small $\Delta_{k}$. Based on this result, we compute the $Im(\omega)$ to the $6^{th}$ WKB order as a function of the hair $\alpha\in(0,0.8)$ as shown in Fig. \ref{imw}. It is noticeable that, for any value of $s$, the $Im(\omega)$ decreases with $\alpha\in(0,\sim0.7)$ which entails a increasing in the damping factor of the signal in comparison with the Schwarzschild case ($\alpha=0$). For $\alpha>0.7$ there is a ``noisy" behaviour $Im(\omega)$ that we attribute to the lack of accuracy of the numerics.

\begin{figure*}[hbt!]
\centering
\includegraphics[width=0.3\textwidth]{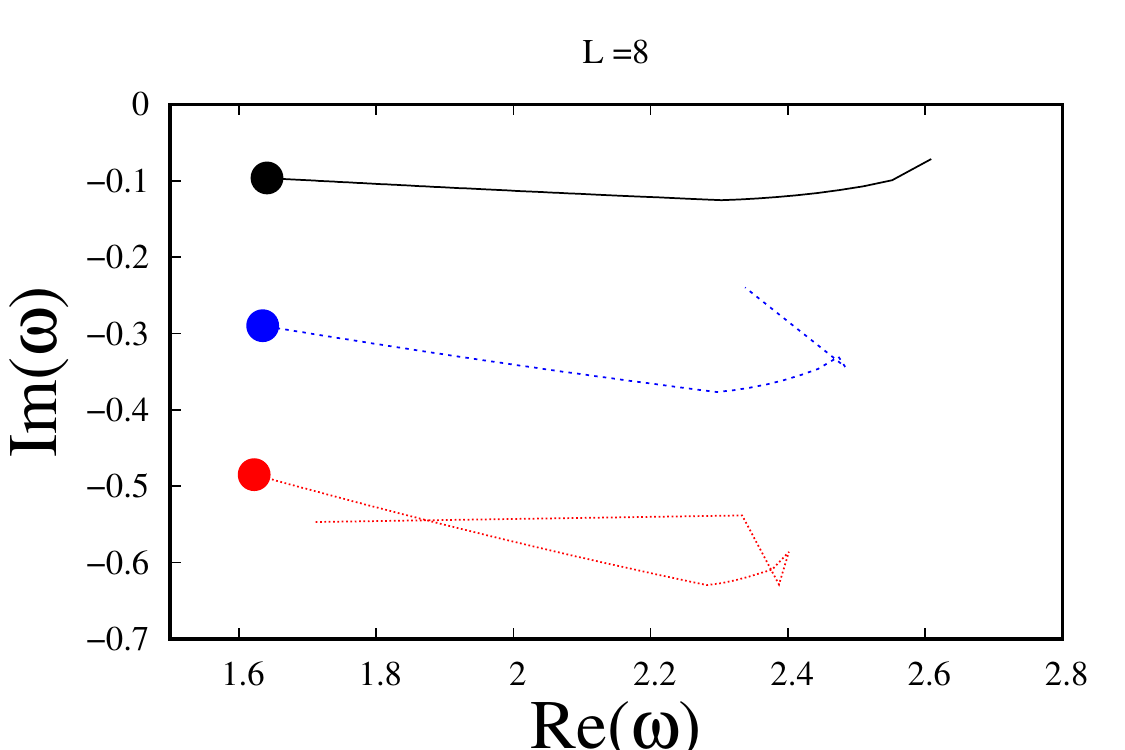}  \
\includegraphics[width=0.3\textwidth]{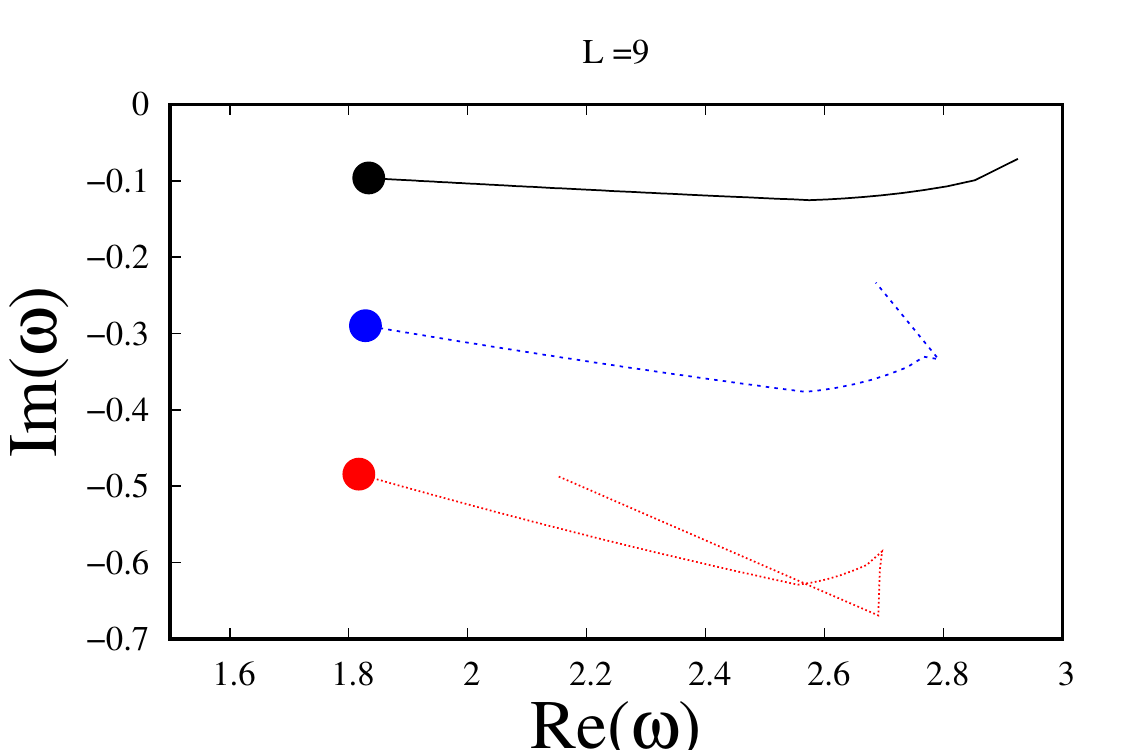}  \
\includegraphics[width=0.3\textwidth]{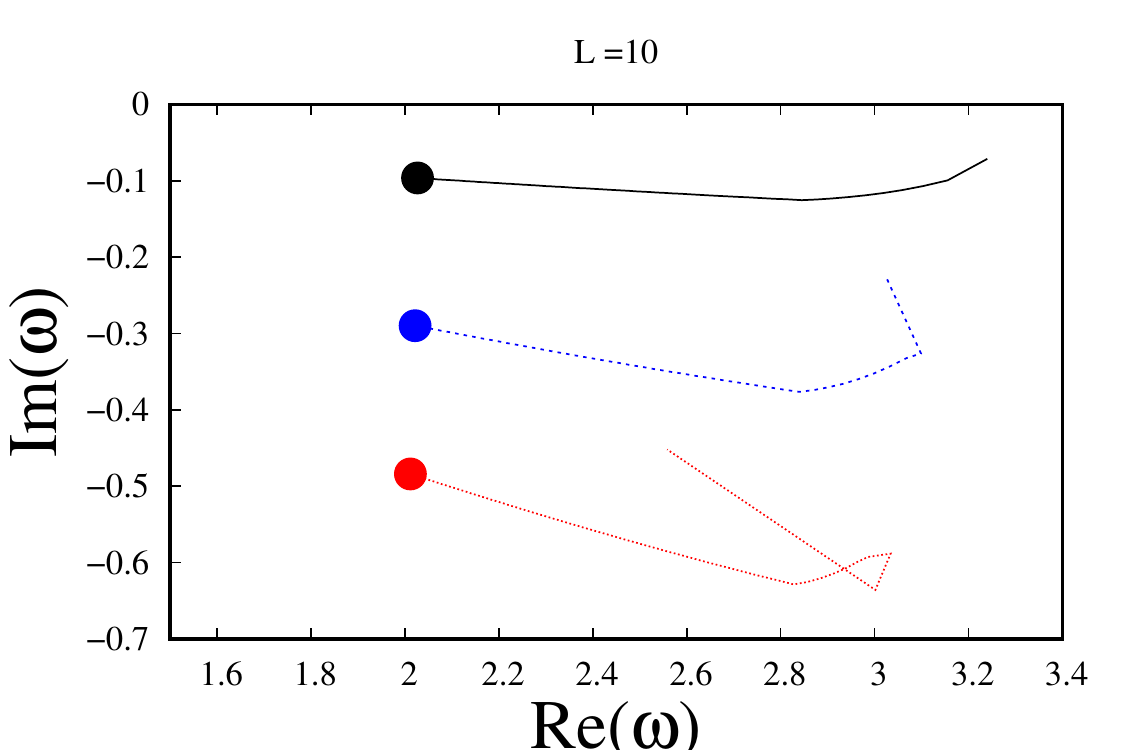}  \
\medskip
\includegraphics[width=0.3\textwidth]{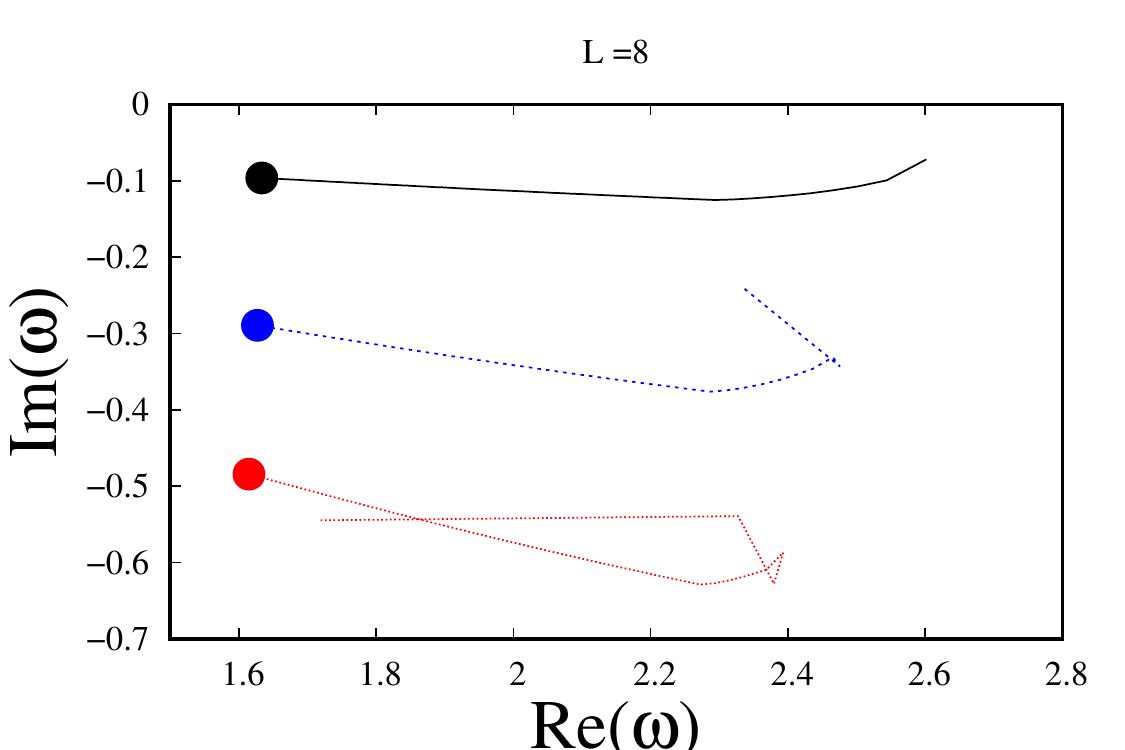}  \
\includegraphics[width=0.3\textwidth]{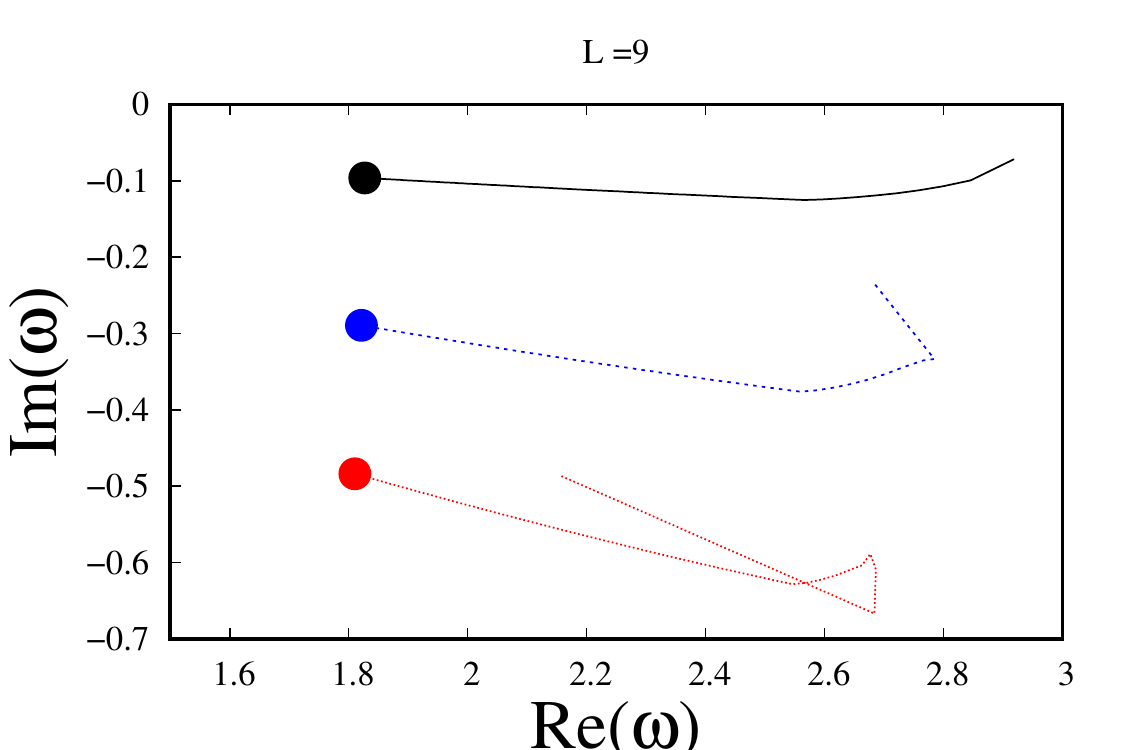}  \
\includegraphics[width=0.3\textwidth]{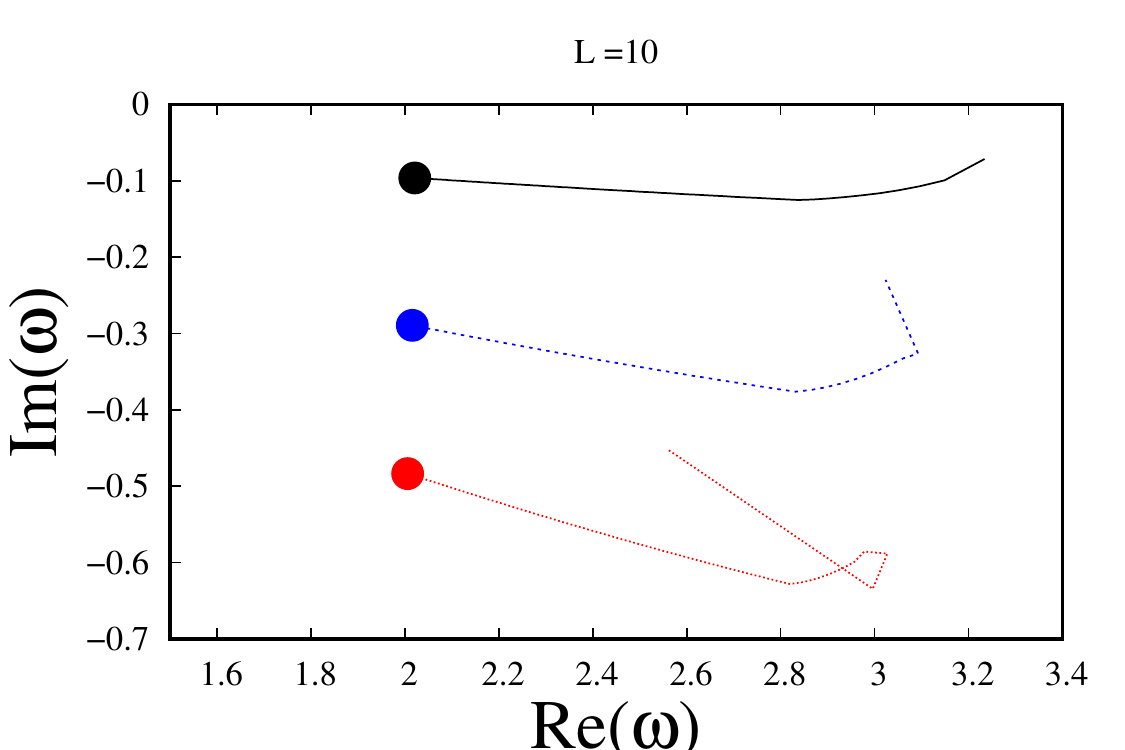}  \
\medskip
\includegraphics[width=0.3\textwidth]{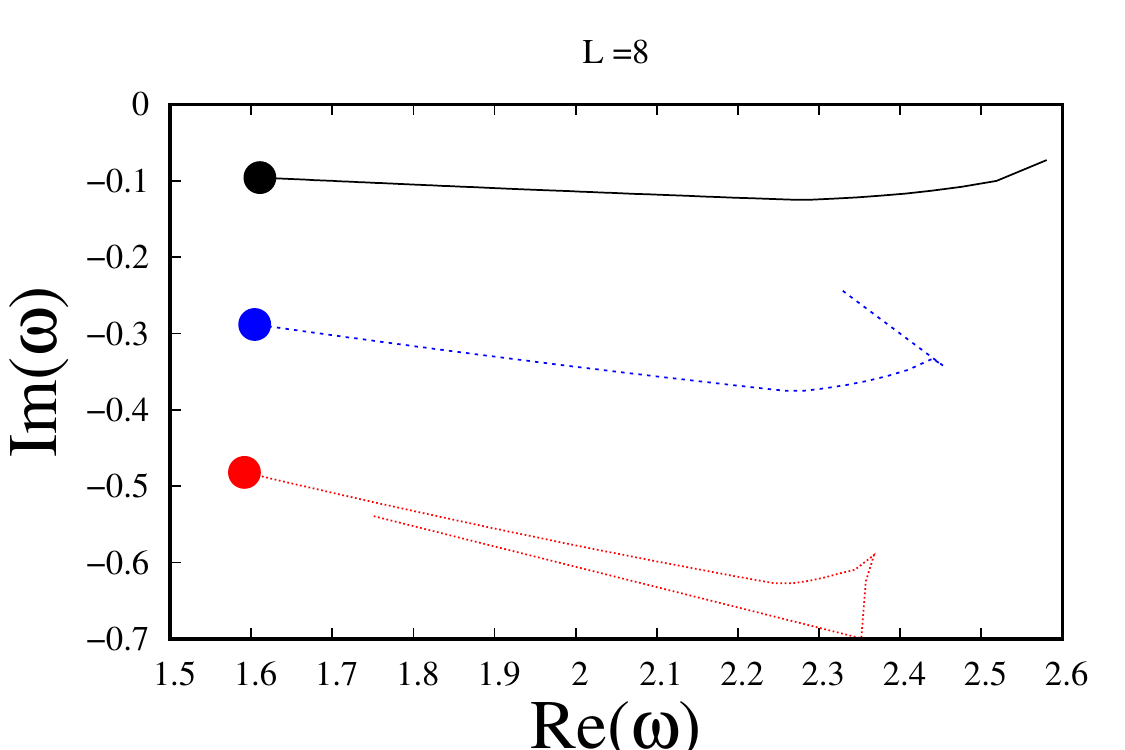}  \
\includegraphics[width=0.3\textwidth]{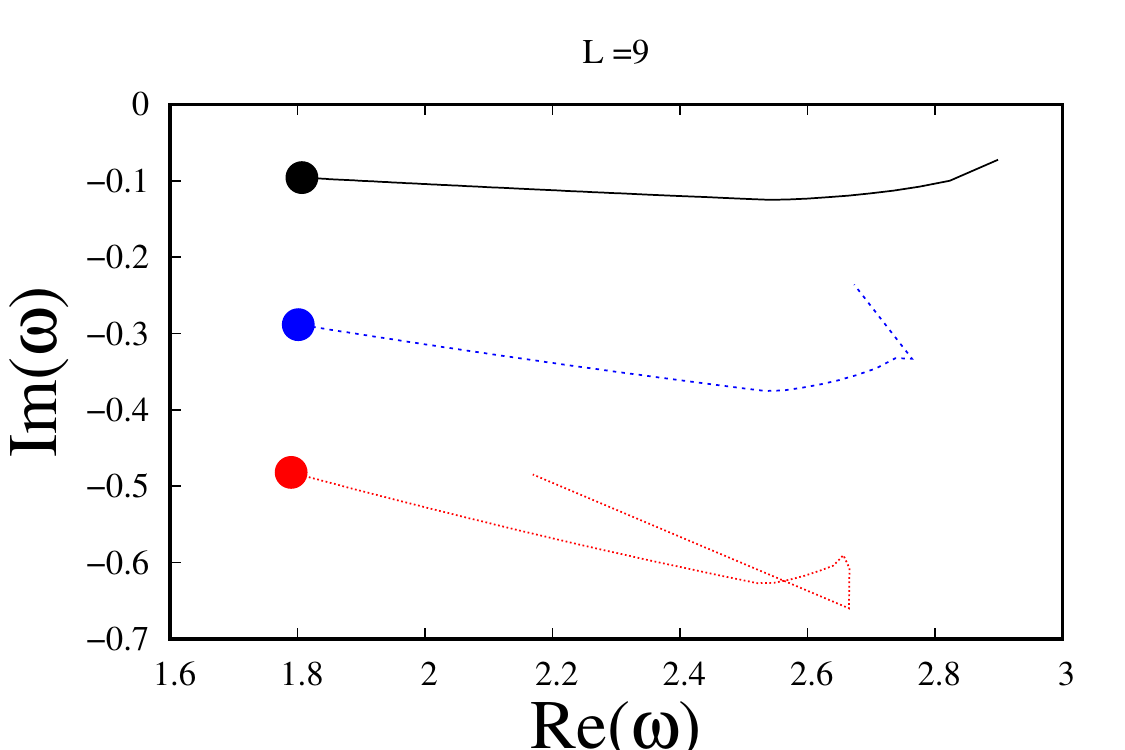}  \
\includegraphics[width=0.3\textwidth]{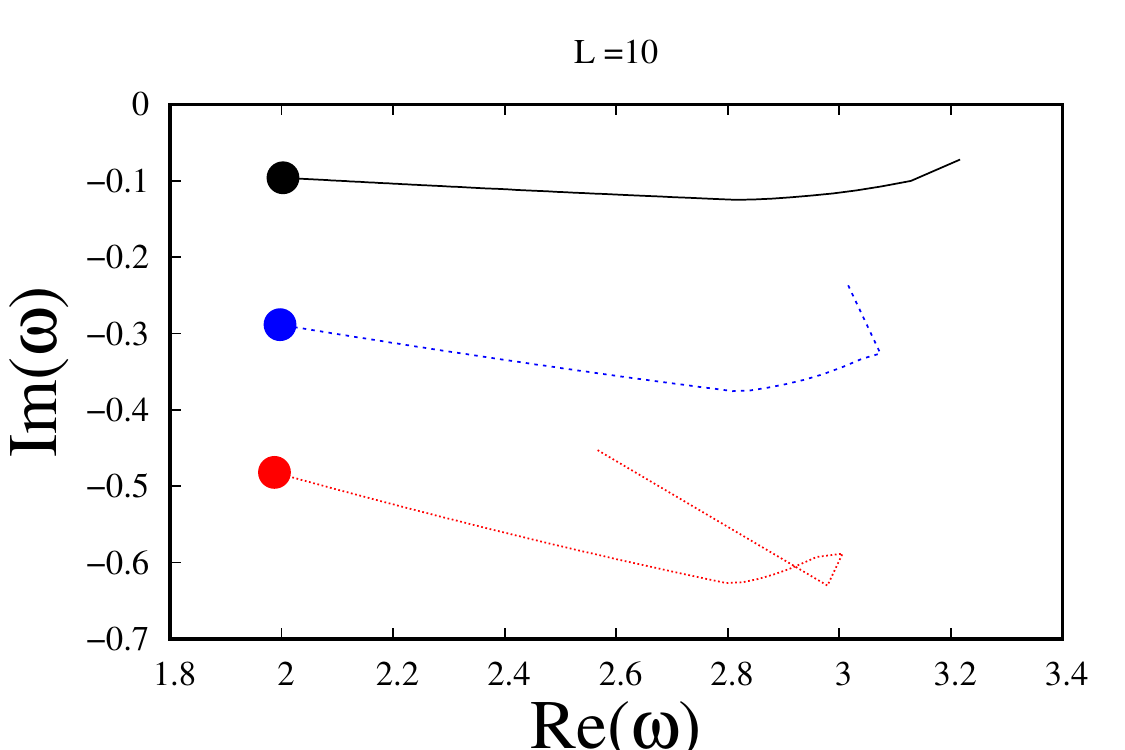}  \
\caption{\label{overtones} $Im(\omega)$ as a function of $Re(\omega)$ for $\ell=8,9,10$ and $n=0$ (black line), $n=1$ (blue line) and $n=2$ (red line) for $s=0$ (frist row), $s=1$ (second row) and $s=2$ (third row). The dots correspond to the Schwarzschild case ($\alpha=0$)}
\end{figure*}

We shall conclude this section by mentioning the role played by the overtones in the analysis of QNM because as it has been shown in \cite{Oshita:2021iyn,Forteza:2021wfq,Jaramillo:2020tuu}, the accurate modeling of the ringdown requires up to ten first overtones  and not only the fundamental mode as is commonly believed \cite{Konoplya:2022hll}. In Fig. (\ref{overtones}) we plot $Im(\omega)$
as a function of $Re(\omega)$ parameterized by the hairy parameter $\alpha$ for scalar (first row), vector (second row) and tensor (third row) perturbations. We note that, in contrast to what occurs with the fundamental mode $n=0$, the behaviour for $n=1$ and $n=2$ present a turning point  for the real part of the QNM frequency. A related but different behaviour is reported in \cite{Konoplya:2022hll} where there is analysed the role of the overtones for a regular black hole in asymptotically safe
gravity.

\begin{table*}[htb!]
\centering
\begin{tabular}{|m{2em}|m{2em}|m{2em}|m{4em}|m{4em}|m{4em}|m{4em}|m{4em}|m{4em}|} 
 \hline
 $s$ & $L$ & $n$ &$\Delta_{2}$ & $\Delta_{3}$ &$\Delta_{4}$ &$\Delta_{5}$&$\Delta_{6}$&$\Delta_{7}$ \\  
\hline
0& 8 & 0 & 0.00378 & 0.00017 & $4\times10^{-6}$ & $5\times10^{-8}$ & \textcolor{red}{$1\times10^{-8}$} & $5\times10^{-8}$\\
& & 1 & 0.01958 & 0.00084 & $4\times10^{-6}$ & $3\times10^{-7}$ & \textcolor{red}{$1\times10^{-7}$} & $9\times10^{-7}$  \\ 
& & 2 & 0.04976 & 0.00244 & 0.00010 & $9\times10^{-6}$ & \textcolor{red}{$8\times10^{-7}$} & $7\times10^{-6}$\\
& 9 & 0 & 0.00338 & 0.00013 & $3\times10^{-6}$ & $3\times10^{-8}$ & \textcolor{red}{$9\times10^{-9}$} & $1\times10^{-8}$ \\
& & 1 & 0.01757 & 0.00067 & $3\times10^{-6}$ & $9\times10^{-8}$ & \textcolor{red}{$8\times10^{-8}$} & $2\times10^{-7}$\\
& & 2 & 0.04492 & 0.00197 & 0.00007 & $6\times10^{-6}$ & \textcolor{red}{$5\times10^{-7}$} & $1\times10^{-6}$ \\
& 10 & 0 & 0.00306 & 0.00011 & $2\times10^{-6}$ & $2\times10^{-8}$ & \textcolor{red}{$5\times10^{-9}$} & $1\times10^{-8}$ \\
& & 1 & 0.01594 & 0.00055 & $2\times10^{-6}$ & $6\times10^{-8}$ & \textcolor{red}{$4\times10^{-8}$} & $3\times10^{-7}$ \\
& & 2 & 0.04091 & 0.00162 & 0.00005 & $4\times10^{-6}$ & \textcolor{red}{$2\times10^{-7}$} & $2\times10^{-6}$ \\
\hline
1& 8 & 0 & 0.00379 & 0.00017 & $4\times10^{-6}$ & $6\times10^{-8}$ & \textcolor{red}{$1\times10^{-8}$} & $6\times10^{-8}$\\
& & 1 & 0.01962 & 0.00084 & $5\times10^{-6}$ & $2\times10^{-7}$ & \textcolor{red}{$1\times10^{-7}$} & $1\times10^{-6}$ \\ 
& & 2 & 0.04987 & 0.00246 & 0.00010 & $9\times10^{-6}$ & \textcolor{red}{$8\times10^{-7}$} & $8\times10^{-6}$\\
& 9 & 0 & 0.00339 & 0.00013 & $3\times10^{-6}$ & $4\times10^{-8}$ & \textcolor{red}{$9\times10^{-9}$} & $1\times10^{-8}$\\
& & 1 & 0.01761 & 0.00067 & $3\times10^{-6}$ & $1\times10^{-7}$ & \textcolor{red}{$8\times10^{-8}$} & $1\times10^{-7}$\\
& & 2 & 0.04500 & 0.00198 & 0.00007 & $6\times10^{-6}$ & \textcolor{red}{$4\times10^{-7}$} & $1\times10^{-6}$\\
& 10 & 0 & 0.00307 & 0.00011 & $2\times10^{-6}$ & $2\times10^{-8}$ & \textcolor{red}{$5\times10^{-9}$} & $8\times10^{-9}$\\
& & 1 & 0.01597 & 0.00055 & $2\times10^{-6}$ & $6\times10^{-8}$ & \textcolor{red}{$5\times10^{-8}$} & $1\times10^{-7}$\\
& & 2 & 0.04097 & 0.00163 & 0.00005 & $4\times10^{-6}$ & \textcolor{red}{$3\times10^{-7}$} & $1\times10^{-6}$\\
\hline
2& 8 & 0 & 0.00381 & 0.00017 & $4\times10^{-6}$ & $7\times10^{-8}$ & \textcolor{red}{$1\times10^{-8}$} & $3\times10^{-8}$\\
& & 1 & 0.01976 & 0.00086 & $5\times10^{-6}$ & $2\times10^{-7}$ & \textcolor{red}{$1\times10^{-7}$} & $4\times10^{-7}$  \\ 
& & 2 & 0.05022 & 0.00253 & 0.00010 & $9\times10^{-6}$ & \textcolor{red}{$8\times10^{-7}$} & $4\times10^{-6}$\\
& 9 & 0 & 0.00341 & 0.00014 & $3\times10^{-6}$ & $4\times10^{-8}$ & \textcolor{red}{$1\times10^{-8}$} & $2\times10^{-8}$ \\
& & 1 & 0.01771 & 0.00069 & $3\times10^{-6}$ & $1\times10^{-7}$ & \textcolor{red}{$8\times10^{-8}$} & $4\times10^{-7}$\\
& & 2 & 0.04525 & 0.00202 & 0.00007 & $6\times10^{-6}$ & \textcolor{red}{$4\times10^{-7}$} & $3\times10^{-6}$ \\
& 10 & 0 & 0.00308 & 0.00011 & $2\times10^{-6}$ & $2\times10^{-8}$ & \textcolor{red}{$6\times10^{-9}$} & $1\times10^{-8}$ \\
& & 1 & 0.01604 & 0.00056 & $2\times10^{-6}$ & $7\times10^{-8}$ & \textcolor{red}{$5\times10^{-8}$} & $2\times10^{-7}$ \\
& & 2 & 0.04117 & 0.00166 & 0.00005 & $4\times10^{-6}$ & \textcolor{red}{$2\times10^{-7}$} & $2\times10^{-6}$ \\
\hline
\end{tabular}
\caption{Numerical values of $\Delta_k$ for $\alpha=0.5$ and different values of $s$, multipole number $L$ and overtone $n$. The values highlighted in red color correspond to the minimum value of $\Delta_{k}$}
\label{table:1}
\end{table*}

\section{Conclusions}
In this work we have constructed a static and spherically symmetric black hole which we interpret as being supported by a family of generic mono-parametric sources thorough Gravitational Decoupling. Interestingly, the parameter which characterizes the matter sector can be interpreted as a genuine hair, 
which cannot be associated to any global charge, after a judicious choice for the aforementioned matter sector. Although the solution was constructed by demanding the weak energy condition, we found that the resulting matter sector satisfies all the energy conditions at and outside the event horizon, which gives stronger support to our choice for the matter sector. Finally, we have studied the effect of the hair on both the periastron advance and the gravitational lensing of the black hole. We have closed our manuscript by estimating the best WKB order to compute the quasinormal frequencies for scalar, vector and tensor perturbation fields.
\subsection*{Acknowledgments} 
P. B. is funded by the Beatriz Galindo contract BEAGAL 18/00207, Spain.

%

%
%

\bibliography{references.bib,referencesQNM.bib}
\bibliographystyle{unsrt}

\end{document}